\documentclass[preprint]{aastex}
\usepackage{amsmath}
\begin{document}

\title{Unveiling the Evolutionary Sequence from Infalling Envelopes to Keplerian Disks around Low-Mass Protostars}
\author{Hsi-Wei Yen\altaffilmark{1,2}, Shigehisa Takakuwa\altaffilmark{2}, Nagayoshi Ohashi\altaffilmark{2,3}, and Paul T. P. Ho\altaffilmark{2,4}}
\altaffiltext{1}{Institute of Astrophysics, National Taiwan University, Taipei 10617, Taiwan; hwyen@asiaa.sinica.edu.tw}
\altaffiltext{2}{Academia Sinica Institute of Astronomy and Astrophysics, P.O. Box 23-141, Taipei 10617, Taiwan} 
\altaffiltext{3}{Subaru Telescope, National Astronomical Observatory of Japan, 650 North A‚ \"A\^oohoku Place, Hilo, HI 96720, USA}
\altaffiltext{4}{Harvard-Smithsonian Center for Astrophysics, 60 Garden Street, Cambridge, MA 02138, USA}

\begin{abstract}
We performed SMA observations in the C$^{18}$O (2--1) emission line toward six Class 0 and I protostars,
to study rotational motions of their surrounding envelopes and circumstellar material on 100 to 1000 AU scales. 
C$^{18}$O (2--1) emission with intensity peaks located at the protostellar positions is detected toward all the six sources. 
The rotational velocities of the protostellar envelopes as a function of radius were measured from the Position--Velocity diagrams perpendicular to the outflow directions passing through the protostellar positions. 
Two Class 0 sources, B335 and NGC 1333 IRAS 4B, show no detectable rotational motion,  
while L1527 IRS (Class 0/I) and L1448-mm (Class 0) exhibit rotational motions with radial profiles of $V_{\rm rot} \propto r^{-1.0\pm0.2}$ and $\propto r^{-1.0\pm0.1}$, respectively. 
The other Class I sources, TMC-1A and L1489 IRS, exhibit the fastest rotational motions among the sample,
and their rotational motions have flatter radial profiles of $V_{\rm rot} \propto r^{-0.6\pm0.1}$ and $\propto r^{-0.5\pm0.1}$, respectively.
The rotational motions with the radial dependence of $\sim r^{-1}$ can be interpreted as rotation with a conserved angular momentum in a dynamically infalling envelope, 
while those with the radial dependence of $\sim r^{-0.5}$ can be interpreted as Keplerian rotation.
These observational results demonstrate categorization of rotational motions from infalling envelopes to Keplerian-disk formation. 
Models of the inside-out collapse where the angular momentum is conserved are discussed and compared with our observational results.
\end{abstract}


\section{Introduction}
Protostars are formed through gravitational collapse of dense cores ($n \sim 10^{4} - 10^{5}$ cm$^{-3}$; e.g., Andr\'{e} et al.~2000; Myers et al.~2000).  
Previous interferometric observations have found infalling and rotational components on thousands of AU scale inside dense cores associated with known infrared sources, so-called protostellar envelopes (e.g., Ohashi et al.~1996, 1997b; Momose et al.~1998). 
In the innermost ($<$ 500 AU) part of those protostellar envelopes, 
circumstellar disks are expected to form around protostars when collapsing material rotates fast enough to reach its Keplerian velocity and to become centrifugally supported (e.g., Shu et al.~1987). 

Around T Tauri stars, 
interferometric observations in millimeter molecular lines have identified such Keplerian disks (e.g., Guilloteau \& Dutrey 1994; Dutrey et al.~1998; Simon et al.~2000; Qi et al.~2003, 2004; Andrews \& Williams 2007; Pietu et al. 2007; Guilloteau et al.~2011; Andrews et al.~2012). 
The radii of those Keplerian disks seen in CO emission range from $\sim 100$ to $\sim 800$ AU, 
and their masses traced by dust continuum emission range from $\sim 10^{-4}$ to 10$^{-1}$ $M_\sun$. 
Recent observations with the Submillimeter Array (SMA)$\footnotemark$ have reported the presence of Keplerian disks around Class I protostars, which are still embedded in protostellar envelopes (Brinch et al.~2007a; Lommen et al.~2008; J{\o}rgensen et al.~2009; Takakuwa et al.~2012). 
The Keplerian disks around Class I protostars have radii from 100 to 300 AU and masses from 0.004 to 0.06 $M_\sun$, comparable to those around T Tauri stars. 

On the other hand, 
circumstellar disks around Class 0 protostars are likely deeply embedded in protostellar envelopes, and difficult to be observed directly (e.g., Looney et al.~2003; Chiang et al.~2008). 
Previous interferometric observations of millimeter dust continuum emission in B335 (Harvey et al.~2003) and L1157-mm (Chiang et al.~2012) at sub-arcsecond angular resolutions have identified possible disk components with the outer radii $<$ 100 AU and $<$ 40 AU, respectively. 
Several previous molecular-line observations of Class 0 sources have inferred the inner centrifugal radii to be $<$ 100 AU, from the comparison between the infalling and  rotational motions of the protostellar envelopes (e.g., Lee et al.~2006, 2009; Yen et al.~2010, 2011). 
Recently, 
Tobin et al.~(2012a) have reported the detection of a possible Keplerian disk with a radius of $\sim 90$ AU around L1527 IRS, a transitional object from the Class 0 to I stages, from their interferometric observations of the $^{13}$CO (2--1) emission, and have claimed that L1527 IRS is the youngest protostar surrounded by a Keplerian disk. 
These observational results combined with those of Class I protostars and T Tauri stars imply that Keplerian disks around protostars likely increase their radii from $< 100$ AU up to $\sim 800$ AU as protostars evolve from Class 0 to T Tauri stages.

Theoretical calculations of gravitational collapse of dense cores where the angular momentum is conserved suggest that the radii of Keplerian disks around protostars increase continuously with evolution (e.g., Ulrich 1976; Cassen \& Moosman 1981; Terebey et al.~1984; Basu 1998). 
As more material falls toward the center of protostellar envelopes with a conserved angular momentum, 
more angular momenta travel to the central region.
Thus the rotational velocities of the material around protostars and the radii of the Keplerian disks increase. 
On the contrary, 
recent magnetohydrodynamic (MHD) simulations show that the magnetic field can effectively remove the angular momentum of the collapsing material by magnetic braking, 
and suppress the growth of the radii of the Keplerian disks within 10 AU (e.g., Mellon \& Li 2008, 2009; Machida et al.~2011; Li et al.~2011; Dapp et al.~2012). 
After the efficiency of magnetic braking decreases due to the dissipation of protostellar envelopes, 
the radii of the Keplerian disks could increase to $\gtrsim$ 100 AU in the end of the main accretion phase. 
However, 
the mass of the formed disk is larger than the protostellar mass by a factor of two to five ($M_{\rm disk} = 0.2 - 1.0\ M_\sun$), 
and the disk is subject to further fragmentation (Machida et al.~2011). 
Such massive disks around Class I protostars, which are close to the end of the main accretion phase, have not been seen observationally. 
It is still unclear as to how Keplerian disks are formed out of protostellar envelopes, increase their radii, and evolve to be classical disks around T Tauri stars.

Formation and evolution of Keplerian disks around protostars should be closely related to the mechanism of angular momentum transfer in collapsing material from envelope (thousands of AU) to disk scales ($\sim 100$ AU). 
Single-dish observations in NH$_3$ and N$_2$H$^+$ emission lines have found that dense cores and protostellar envelopes exhibit velocity gradients over a 0.1 pc scale with a mean magnitude of $\sim 1 - 2$ km s$^{-1}$ pc $^{-1}$, suggestive of large-scale rotational motions (Goodman et al.~1993; Gaselli et al.~2002; Tobin et al.~2011). 
Interferometric N$_2$H$^+$ observations have shown that the inner part of protostellar envelopes (thousands of AU scale) exhibits a larger amount of velocity gradients with a mean magnitude of $\sim7-8$ km s$^{-1}$ pc $^{-1}$ (Chen et al.~2007; Tobin et al.~2011), 
suggesting that the inner envelopes rotate faster than the outer envelopes. 
On the other hand, 
CARMA N$_2$H$^+$ observations of L1157-mm (Chiang et al.~2010) and Very Large Array (VLA) NH$_3$ observations of HH 211 (Tanner \& Arce 2011) have found that the protostellar envelopes on thousands of AU scale around these two Class 0 sources exhibit rigid-body rotation with the magnitudes of 1.5 and 6.3 km s$^{-1}$ pc$^{-1}$, respectively. 
In reality, 
the overall velocity gradients seen in the protostellar envelopes should reflect combination of rotational motion and other systematic gas motions such as infalling and/or outflowing motions, as well as the envelope morphologies (Tobin et al.~2012b),  
and it is not straightforward to disentangle these different motions and envelope morphologies and to extract rotational components.  
Higher angular-resolution interferometric observations of the innermost parts of protostellar envelopes on hundreds of AU scale are desirable to extract rotational components, 
because the rotational velocities are expected to be larger in innermost envelopes if the angular momentum is conserved in collapsing envelopes.

To study formation and evolution of Keplerian disks inside protostellar envelopes, 
it is crucial to observationally identify rotational motions of protostellar envelopes and their radial profiles from envelope to disk scales. 
In previous observational studies of protostellar envelopes (e.g., Ohashi et al.~1997a; Momose et al.~1998; Takakuwa et al.~2004; Yen et al.~2011), 
the radial profiles of rotational velocities have been presumed to be either $V_{\rm rot} \propto r^{-1}$ (rotation conserving its angular momentum) or $V_{\rm rot} \propto r^{-0.5}$ (Keplerian rotation). 
On the other hand,  
previous claims of detection of Keplerian disks around Class I protostars are based on the presence of clear velocity gradients and elongated structures perpendicular to the outflow directions and/or the results of fitting envelope + disk models to the visibility data in millimeter continuum and molecular-line emission (e.g., Brinch et al.~2007a; Lommen et al.~2008; J{\o}rgensen et al.~2009). 
In these studies of Keplerian disks, 
rotational motions have been presumed to follow Keplerian rotation, 
and the radial profiles of rotational velocities were not directly measured from the data. 
Systematical and unambiguous comparison of radial profiles of angular momenta on 100 to 1000 AU scales among protostellar sources at different evolutionary stages has not been made. 
In order to directly measure the radial distributions of angular momenta in protostellar sources without any presumption of radial profiles of rotational velocities, 
we have developed our own analytic method to measure the power-law indices of the rotational profiles. 
In this paper, 
we have conducted SMA observations in the C$^{18}$O (2--1; 219.560358 GHz) emission line toward six protostars at different evolutionary stages from Class 0 to I, 
and measured the rotational velocities of their protostellar envelopes as a function of radius on 100 to 1000 AU scales.  
We have identified three types of the rotational motions on 100 to 1000 AU scales, depending on the protostellar evolutionary stages. 
We discuss and compare our observational results with models of the inside-out collapse in the context of Keplerian-disk formation (Ulrich 1976; Cassen \& Moosman 1981; Terebey et al.~1984; Basu 1998).  

\section{Sample}
\subsection{Overview}
The target sources in this project were selected from the sample of the SMA survey project, PROSAC (J{\o}rgensen et al.~2007, 2009).  
The PROSAC sample was chosen based on the large single-dish survey of submillimeter continuum and molecular-line emission toward low-mass protostars (J{\o}rgensen et al.~2002, 2004a).  
We selected all the nearby ($d$ $<$ 250 pc) sources in the northern hemisphere from the PROSAC sample, except for NGC 1333 IRAS 2A and 4A, 
which have rather complex outflow morphologies (J{\o}rgensen et al.~2004b, 2007). 
Table \ref{sample} presents the summary of our selected sources. 
The selected sources are surrounded by protostellar envelopes (Hogerheijde et al.~1998; Motte \& Andr\'{e} 2001), have bipolar molecular outflows (Hogerheijde et al.~1998; J{\o}rgensen et al.~2007), and exhibit signs of rotational motions on thousands of AU scale (Ohashi et al.~1997a, b; Curiel et al.~1999; Saito et al.~1999; Hogerheijde 2001) except NGC 1333 IRAS 4B. 
Hence, these are excellent targets for the study of rotational motions of the inner envelopes on 100 to 1000 AU scales. 

Our sample contains three Class 0 (B335, NGC 1333 IRAS 4B, and L1448-mm), one Class 0/I (L1527 IRS), and two Class I (TMC-1A, and L1489 IRS) sources.
The 1.3 mm continuum peak positions of these sources, as observed with the SMA, are adopted as the positions of the protostars (J{\o}rgensen et al.~2007, 2009; Brinch et al.~2007a). 
All the velocities shown in this paper are with respect to the systemic velocity ($V_{\rm sys}$) of each source.  
Adopted $V_{\rm sys}$ in the LSR frame are shown in Table \ref{sample}. 
Inferred inclination angles of the envelope/disk mid-plane from the plane of the sky and position angles of the outflow axes of these sources are listed in Table \ref{sample}. 

\footnotetext{The Submillimeter Array (SMA) is a joint project between the Smithsonian Astrophysical Observatory and the Academia Sinica Institute of Astronomy and Astrophysics and is funded by the Smithsonian Institute and the Academia Sinica.}

\subsection{Individual Sources} 
{\it B335}: This source is an isolated Bok globule associated with a Class 0 protostar having a bolometric luminosity of 1.5 $L_{\sun}$ (IRAS 19347+0727; Keene et al.~1980, 1983) at a distance of 150 pc (Stutz et al.~2008). 
B335 is associated with an east-west elongated, conical-shaped molecular outflow, 
and the inclination angle of the disk plane is estimated to be $\sim 80\degr$ (Hirano et al.~1988). 
Along the outflow axis there are also collimated $^{12}$CO (2--1) jets (Yen et al.~2010) and Herbig-Haro objects (HH 119 A--F; Reipurth et al.~1992; G{\aa}lfalk \& Olofsson 2007).
Millimeter interferometric observations in the C$^{18}$O (2--1) and H$^{13}$CO$^{+}$ (1--0) emission lines have revealed the infalling motion of the envelope around B335, 
and the protostellar mass is estimated to be $\sim$ 0.04 $M_\sun$ from the infalling velocity (Saito et al.~1999; Yen et al.~2010). 
The rotational motion of the envelope around B335 has been detected at radii from $\sim20,000$ to $\sim1000$ AU (Saito et al.~1999; Yen et al.~2011), 
while the previous SMA observations did not detect clear rotational motion on hundreds of AU scale, 
and the upper limit of the rotational velocity is estimated to 0.04 km s$^{-1}$ at a radius of 370 AU (Yen et al.~2010). 
The radius of the central circumstellar disk has been estimated to be less than 100 AU from the interferometric observations of the millimeter continuum emission in B335 (Harvey et al.~2003). 
$V_{\rm sys}$ of B335 is 8.3 km $^{-1}$ as measured by the single-dish observations in the C$^{18}$O (2--1; 1--0) emission (Saito et al.~1999; Yen et al.~2011).

{\it NGC 1333 IRAS 4B}: This source is a Class 0 protobinary system with a bolometric luminosity of 1.6 $L_\sun$ (Enoch et al.~2009b) in the Perseus molecular cloud ($d$ = 250 pc; Enoch et al.~2006).
The separation of the binary companions, named 4B and 4B$^\prime$, is $\sim2750$ AU ($\sim11\arcsec$; J{\o}rgensen et al.~2007). 
In the present paper, 
we refer the primary component 4B to as IRAS 4B and discuss the molecular emission associated with the primary component. 
IRAS 4B is associated with a bipolar $^{12}$CO outflow along the north-south direction (e.g., J{\o}rgensen et al.~2007) as well as a collimated jet seen in the H$_{2}$O maser emission (e.g., Marvel et al.~2008). 
The position angle of the H$_2$O jet is different from that of the $^{12}$CO outflow by $\sim30\degr$.  
From the proper motions of the H$_{2}$O maser spots, the inclination angle of the disk plane of IRAS 4B is estimated to be $\sim77\degr$ (Marvel et al.~2008). 
Interferometric observations of the H$_2$CO (3$_{12}$--2$_{11}$) emission in IRAS 4B show an inverse P Cygni profile, suggesting the presence of infalling motion (Di Francesco et al.~2001). 
$V_{\rm sys}$ of IRAS 4B is 6.7 km s$^{-1}$ as measured from the combined BIMA+FCRAO C$^{18}$O (1--0) spectrum (Volgenau et al.~2006). 
The NGC 1333 IRAS 4 region is complex and contains a small cluster of protostars within a radius of $\sim30\arcsec$ (7500 AU), 
and no systematic velocity gradient corresponding to rotational motion on thousands of AU scale was observed (Volgenau et al.~2006). 

{\it L1527 IRS} (IRAS 04368+2557): 
This source is a protostar with a bolometric luminosity of 2.8 $L_\sun$ (Tobin et al.~2008) in the Taurus molecular cloud ($d$ = 140 pc). 
The source exhibits a Class 0 type spectrum, but could be an obscured Class I protostar because of its high inclination angle (Ohashi et al.~1997a). 
The inclination angle of the disk plane is estimated to be 85$\degr$ from the infrared images (Tobin et al. 2008).
Previous millimeter interferometric observations show that the $^{12}$CO (1--0) outflow associated with L1527 IRS exhibits a bipolar conical shape along the east-west direction,  
and that there are infalling and rotational motions in the C$^{18}$O (1--0) flattened envelope on a 2000 AU scale (Ohashi et al.~1997a). 
From the mass infalling rate ($\sim1 \times 10^{-6}\ M_\sun$ yr$^{-1}$) estimated by Ohashi et al.~(1997a) and the bolometric luminosity (2.8 $L_\sun$) measured by Tobin et al.~(2008), 
the protostellar mass is estimated to be 0.2 $M_\sun$. 
Recently the presence of a possible Keplerian disk with a radius of 90 AU and a mass of 0.007 $M_\sun$ around L1527 IRS has been reported from the CARMA observations in the $^{13}$CO (2--1) emission line,  
and the protostellar mass has been estimated to be 0.2 $M_\sun$ from the observed Keplerian rotation (Tobin et al.~2012a). 
$V_{\rm sys}$ of L1527 IRS is measured to be 5.7 km s$^{-1}$ from the Nobeyama 45-m telescope observations in the C$^{18}$O (1--0) emission line (Ohashi et al.~1997a),
while that is measured to be 5.9 km s$^{-1}$ by the N$_2$H$^+$ and NH$_3$ single-dish observations (Goodman et al.~1993; Caselli et al.~2002; Tobin et al~2011). 
As will be presented below, 
our SMA results show that the distribution of the C$^{18}$O (2--1) emission is symmetric with respect to $V_{\rm LSR}$ = 5.7 km s$^{-1}$, 
and hence $V_{\rm sys}$ of 5.7 km s$^{-1}$ is adopted in this paper.
 
{\it L1448-mm}: This source is also known as L1448 C, a Class 0 protostar with a bolometric luminosity of 7.5 $L_\sun$ (Tobin et al.~2007) in the Perseus molecular cloud. 
L1448-mm is associated with an active molecular outflow along the northwest-southeast direction, and 
several high-velocity ($>$ 50 km s$^{-1}$) $^{12}$CO and H$_2$O "bullets" along the outflow axis and collimated SiO jets have been observed (e.g., Bachiller 1996; Hirano et al.~2010; Kristensen et al.~2011).
From the proper motions of the SiO jets, 
the inclination angle of the disk plane is estimated to be $\sim 70\degr$ (Girart \& Acord 2001). 
Based on the mass loss rate of the outflow and the bolometric luminosity, 
the protostellar mass of L1448-mm is inferred to be 0.03 -- 0.09 $M_\sun$ (Hirano et al.~2010).
VLA observations in the NH$_{3}$ emission lines of this source show that the envelope around L1448-mm exhibits velocity gradients both along and perpendicular to the outflow direction, 
which are interpreted as infalling and rotational motions on thousands of AU scale, respectively (Curiel et al.~1999).  
$V_{\rm sys}$ of L1448-mm is 5.0 km s$^{-1}$ as measured by the VLA NH$_{3}$ observation (Curiel et al.~1999).

{\it TMC-1A} (IRAS 04365+2535): This source is a Class I protostar with a bolometric luminosity of 2.4 $L_\sun$ (Furlan et al.~2008) in the Taurus molecular cloud. 
TMC-1A is associated with a conical-shaped, bipolar molecular outflow along the northwest-southeast direction, 
and the inclination angle of the disk plane is estimated to be $40\degr - 70\degr$ (Chandler et al.~1996). 
Interferometric observations of the C$^{18}$O (1--0) emission in TMC-1A (Ohashi et al.~1997b) show that the envelope exhibits a velocity gradient perpendicular to the outflow direction only, suggesting that the rotational motion is more dominant than the infalling motion on a 1500 AU scale. 
$V_{\rm sys}$ of TMC-1A is 6.6 km s$^{-1}$ as estimated by the IRAM 30-m telescope and the James Clerk Maxwell Telescope (JCMT) observations in the C$^{18}$O and $^{12}$CO emission lines at multiple transitions (Hogerheijde et al.~1998). 

{\it L1489 IRS} (IRAS 04016+2610): This source is a Class I protostar with a bolometric luminosity of 3.7 $L_\sun$ (Furlan et al.~2008) in the Taurus molecular cloud. 
In L1489 IRS, 
there is a faint molecular outflow along the northwest-southeast direction as observed by the JCMT in the $^{12}$CO (3--2) emission line (Hogerheijde et al.~1998). 
Estimates of the inclination angle of the disk plane range from 36$\degr$ to 90$\degr$ from the model fitting of the spectral energy distribution, scatter light and millimeter continuum images, and molecular-line spectra (Kenyon et al.~1993; Padgett et al.~1999; Eisner et al.~2005; Brinch et al.~2007a; Eisner 2012). 
The infalling and rotational motions of the envelope on a 2000 AU scale have been revealed by interferometric observations at millimeter wavelengths, 
and the rotational motion is more dominant than the infalling motion (Hogerheijde 2001). 
The SMA observations in the millimeter continuum and HCO$^{+}$ (3--2) line emission have reported the presence of a Keplerian disk with a radius of 200 AU and a mass of 0.004 $M_\sun$ embedded in the envelope, 
and the protostellar mass is estimated to be 1.4 $M_\sun$ on the assumption that the inclination angle of the Keplerian disk is 40$\degr$ (Brinch et al.~2007a). 
The 1.3 mm continuum observations with the CARMA at a sub-arcsecond angular resolution ($\sim0\farcs8$) have shown that the disk radius is 250 -- 450 AU and the disk mass is 0.005 $M_\sun$ (Eisner 2012).
$V_{\rm sys}$ of L1489 IRS is 7.2 km s$^{-1}$ as estimated by the single-dish observations in several molecular lines at multiple transitions (Brinch et al.~2007b).

\section{Observations} 
The data in this work were taken from both the SMA data archive and new SMA observations.
Details of the SMA are described by Ho et al.~(2004).
We conducted SMA observations in the C$^{18}$O (2--1) line with the compact configuration toward TMC-1A and L1489 IRS and with the subcompact configuration toward all the sample sources except B335.  
The SMA C$^{18}$O (2--1) data of IRAS 4B with the extended configuration are obtained from the SMA data archive,   
and the PI of this observation is J{\o}rgensen, J. K..
Details of these observations are shown in Table \ref{obsummary}. 
The SMA C$^{18}$O (2--1) data of B335, IRAS 4B, L1448-mm, and L1527 IRS with the compact configuration are obtained from the SMA data archive,
and the details of these observations are described in the PROSAC paper (J{\o}rgensen et al.~2007). 
The 1.3 mm continuum emission was observed simultaneously in all of these observations.
The maximum and minimum $uv$ lengths after the compilation of all the available data are listed in Table \ref{resolution}. 
In the observations with the compact and extended configurations, 
the correlator configuration was set to assign 512 channels to one chunk with a 83.3 MHz bandwidth for the C$^{18}$O (2--1) line, 
resulting in a velocity resolution of 0.28 km s$^{-1}$.   
In the observations with the subcompact configuration, 1024 channels were assigned to the chunk for the C$^{18}$O (2--1) line, 
resulting in a velocity resolution of 0.14 km s$^{-1}$. 
MIR software package (Scoville et al.~1993) was used to calibrate all the data. 
The calibrated visibility data were Fourier-transformed and CLEANed with MIRIAD (Sault et al.~1995) to produce images.
Two types of images with different velocity and angular resolutions were produced: images made only with the subcompact data and combined images made with all the available data. 
The resolutions and noise levels of the images are summarized in Table \ref{resolution}.
The combined images with the higher angular resolutions can show finer structures, 
while the subcompact images with more weights on the short-spacing data and the higher velocity resolution can trace more extended structures and slower motions. 
In the present paper results of the C$^{18}$O (2--1) data are primarily discussed, 
and detailed discussion of the 1.3 mm continuum data will be the subject to the forthcoming papers. 
The continuum images are shown in Appendix A.

\section{Spatial and Kinematics Structures of the C$^{18}$O (2--1) Emission}

Figure \ref{moment} shows integrated-intensity (i.e., moment 0) maps (contours) overlaid on the intensity-weighted mean velocity (i.e., moment 1) maps (color)  of all the sources in the C$^{18}$O (2--1) emission made from the combined (left) and subcompact data (right).
In the moment 1 maps a common color scale is adopted to compare the amount of the velocity gradients among the different sources directly. 
In all the sources the C$^{18}$O (2--1) emission shows $\sim$ a few thousand AU scale envelope features approximately centered on the protostellar sources, 
but the velocity structures are remarkably different as described below.

\subsection{B335} 
\begin{figure}
\epsscale{0.75}
\plotone{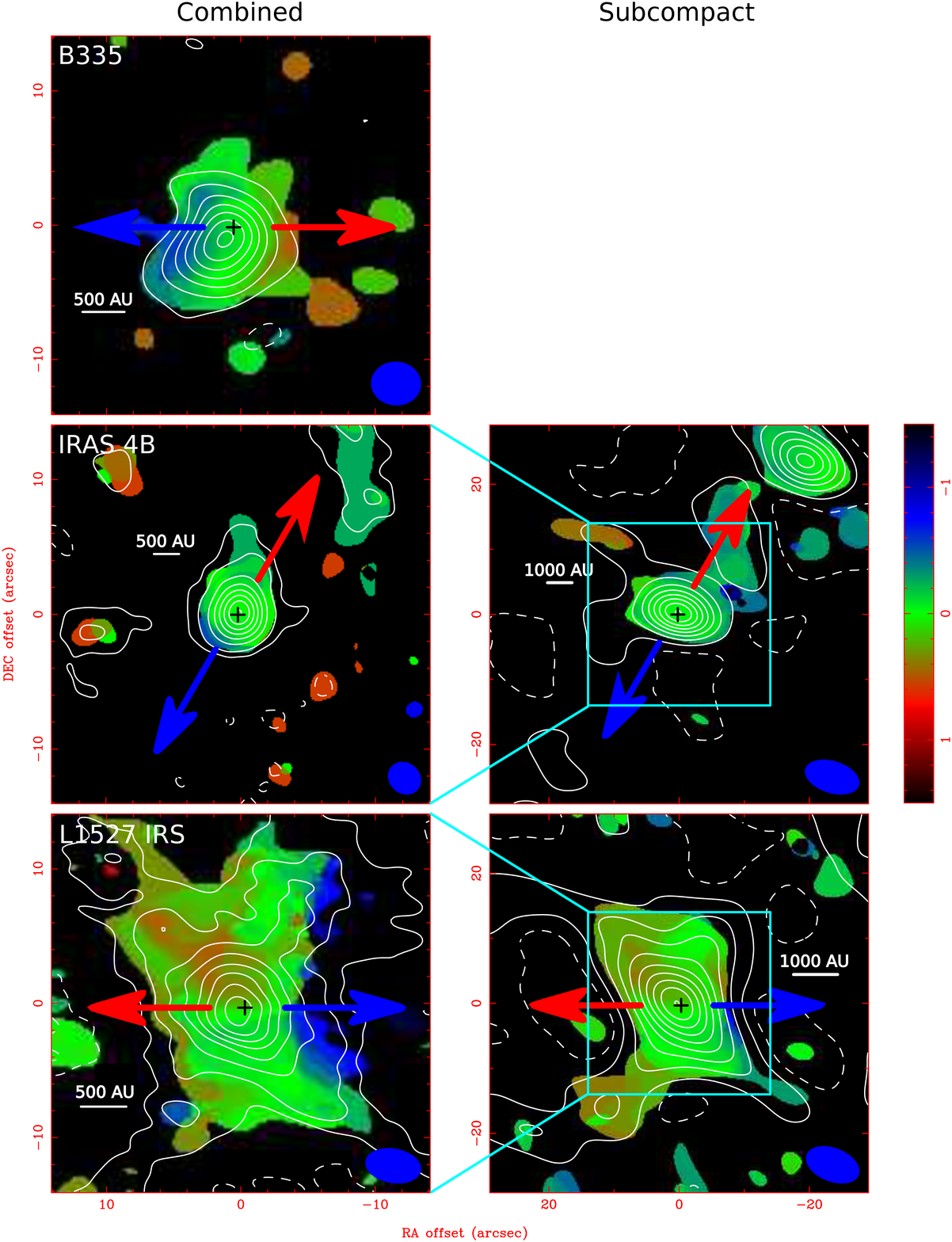}
\caption{Moment 0 maps (contours) overlaid on the moment 1 maps (color scale) of the C$^{18}$O (2--1) emission in B335, IRAS 4B, L1527 IRS, L1448-mm, TMC-1A, and L1489 IRS. The maps in the left and right columns are produced with the combined and subcompact data, respectively. Light blue boxes in the subcompact maps show the area of the combined maps. Crosses present the protostellar positions derived from the 1.3 mm continuum emission (J{\o}rgensen et al.~2007, 2009; Brinch et al.~2007a), and blue and red arrows show the directions of the blueshifted and redshifted outflows, respectively. A blue filled ellipse at the bottom-right corner in each panel denotes the beam size. Contour levels start from 3$\sigma$ in steps of 3$\sigma$ in the combined maps and in steps of 7$\sigma$ in the subcompact maps. The 1$\sigma$ noise levels of the combined moment 0 maps of B335, IRAS 4B, L1527 IRS, L1448-mm, TMC-1A, and L1489 IRS are 0.14, 0.07, 0.1, 0.09, 0.16, and 0.18 Jy beam$^{-1}$ km s$^{-1}$, respectively, and those of the subcompact moment 0 maps of IRAS 4B, L1527 IRS, L1448-mm, TMC-1A, and L1489 IRS are 0.08, 0.12, 0.1, 0.12, and 0.12 Jy beam$^{-1}$ km s$^{-1}$, respectively. The beam sizes are summarized in Table \ref{resolution}.}\label{moment}
\end{figure}

The SMA results of B335 in the C$^{18}$O (2--1) emission have been presented in detail by Yen et al.~(2010, 2011). 
The C$^{18}$O (2--1) emission shows a compact ($\sim1500$ AU) blob approximately centered on the protostar (Fig.~\ref{moment}). 
A velocity gradient along the outflow axis in the east-west direction is clearly seen in the moment 1 map (Fig.~\ref{moment}). 
The velocity gradient of the C$^{18}$O (2--1) emission has been compared with that of the associated outflow traced by the $^{12}$CO (2--1) emission, 
and is unlikely originated from the outflow. 
Through the comparison between the observed C$^{18}$O (2--1) results and models of infalling envelopes,  
we have found that the velocity gradient of the C$^{18}$O (2--1) emission can be explained by the infalling motion in the flattened envelope (see Yen et al.~2010 for details).
On the other hand,  
there is no clear velocity gradient perpendicular to the outflow direction, 
and hence the rotational velocity of the envelope is likely below the detection limit of our SMA observations (Yen et al.~2010, 2011).

\subsection{IRAS 4B}
In the combined moment 0 map of IRAS 4B (Fig.~\ref{moment}), the C$^{18}$O (2--1) emission shows a compact ($\sim$ 2000 AU) condensation centered on the protostar, plus an extension toward the direction of the redshifted outflow. 
The relevant moment 1 map of of IRAS 4B shows that the southeastern part of the condensation is slightly ($\sim$ 0.3 km s$^{-1}$) blueshifted than the northwestern part. 
The direction and the sense of the velocity gradient are the same as those of the associated molecular outflow. 
This velocity gradient is more clearly seen in the velocity channel maps shown in Figure \ref{iras4B}. 
At velocities from $-0.81$ to $-0.53$ km s$^{-1}$, the centroid position of the C$^{18}$O (2--1) emission is shifted to southeast from the protostellar position, while at velocities from 0.02 to 0.58 km s$^{-1}$ the emission peak appears to be located to the northwest. 
The origin of this velocity gradient could be either infalling motion in the flattened envelope as in the case of B335 (e.g., Yen et al. 2010, 2011), or the outflow itself. 
On the other hand, there is no detectable velocity gradient perpendicular to the outflow direction.

The northwestern extension appears in the blueshifted velocities ($-0.53$ to $-0.26$ km s$^{-1}$) and is coincident with the redshifted outflow. 
This component may trace a shell of the northwestern outflow, whose axis is close to the plane of the sky ($\sim13\degr$). 
In the moment 0 map of the subcompact data, there is a secondary component toward the northwest, which is associated with NGC 1333 IRAS 4A. 
In the moment 0 map of the combined data, toward the east there appears a weak component associated with 4B$^\prime$.

\subsection{L1527 IRS}
In the moment 0 maps of L1527 IRS (Fig.~\ref{moment}),  
the C$^{18}$O (2--1) emission shows a central component plus an extended structure delineating the conical shell of the outflow. 
The velocity channel maps of the combined data shown in Figure \ref{L1527} also exhibits a central component plus an X-shaped structure (e.g., at velocities of $-0.23$, 0.05, and 0.88 km s$^{-1}$).  
A similar morphological feature has also been observed in the $^{13}$CO (1--0) emission with the Nobeyama Millimeter Array (Ohashi et al.~1997a), 
which is interpreted as the protostellar envelope plus the outflow shell (see Figure 7 in Ohashi et al.~1997a). 
The peak position of the central component shifts from south to north at blueshifted to redshifted velocities,   
suggesting that there is a velocity gradient perpendicular to the outflow direction. 
This velocity gradient perpendicular to the outflow direction is also seen in the moment 1 maps (Fig.~\ref{moment}). 
The same sense of the velocity gradient has also been identified in interferometric observations of L1527 IRS in the C$^{18}$O (1--0) emission line at an angular resolution of $\sim$ 6$\arcsec$ (Ohashi et al.~1997a) and in the $^{13}$CO (2--1) emission line at an angular resolution of $\sim$ 1$\arcsec$ (Tobin et al.~2012a).
The extended X-shaped structure exhibits a velocity gradient consistent with that of the large-scale outflow (e.g., Ohashi et al.~1997a).  

\subsection{L1448-mm} 
In the C$^{18}$O (2--1) velocity channel maps of the combined data of L1448-mm (Fig.~\ref{L1448}), 
there appears a central compact ($\sim$ 1500 AU) component associated with the protostar as well as an extended ($\sim$ 5000 AU) component elongated along the outflow direction from northwest to southeast, 
plus a secondary component at velocities from 0.52 to 1.07 km s$^{-1}$. 
The extended and secondary components are most likely related to the powerful molecular outflow driven from L1448-mm (e.g., Bachiller 1996; Hirano et al.~2010; Kristensen et al.~2011). 
On the other hand, 
the central compact emission associated with L1448-mm exhibits a velocity gradient along northeast (blueshifted) to southwest (redshifted),  
suggesting a velocity gradient perpendicular to the outflow direction. 
The moment 0 maps of the C$^{18}$O (2--1) emission in L1448-mm (Fig.~\ref{moment}) primarily shows the extended component elongated along the outflow direction,   
and the central component seen in the velocity channel maps appears embedded in the extended component.  
As a result, the velocity gradient of the central compact component is not clear in the moment 1 maps. 
The moment 1 maps show that the southern part of the emission is redshifted, 
which arises from the secondary component seen in the velocity channel maps. 

\subsection{TMC-1A} 
Figure \ref{TMC1A} presents the velocity channel maps of the combined data in TMC-1A. 
The peak positions of the C$^{18}$O (2--1) emission in TMC-1A are shifted from northeast to southwest of the protostar from blueshifted to redshifted velocities, 
suggesting the presence of a velocity gradient perpendicular to the outflow direction. 
The high-velocity components ($|V| \gtrsim 1.4$ km s$^{-1}$) are more compact and closer to the protostar than the low-velocity components ($|V| \lesssim 0.8$ km s$^{-1}$).
In the moment 0 maps (Fig.~\ref{moment}), 
the peak position of the total integrated C$^{18}$O (2--1) emission is coincident with the position of the protostar. 
The velocity gradient perpendicular to the outflow direction seen in the velocity channel maps is clearly identified in the moment 1 maps.   

\subsection{L1489 IRS}
Figure \ref{L1489} presents the velocity channel maps of the combined data in L1489 IRS. 
In the blueshifted velocities the C$^{18}$O (2--1) emission is located to the northeast of the protostar, 
while in the redshifted velocities to the southwest. 
Thus, there is a clear velocity gradient perpendicular to the outflow direction. 
The moment 0 and I maps show an elongation and a clear velocity gradient perpendicular to the outflow direction (Fig.~\ref{moment}), respectively, 
and the peak position of the moment 0 maps is coincident with the position of the protostar. 

In summary, 
the C$^{18}$O (2--1) emission in these sources observed with the SMA shows central condensations associated with the protostars plus extensions toward the outflow directions, 
and most likely traces the protostellar envelopes with the contaminations from the outflows. 
Two Class 0 sources, B335 and IRAS 4B, exhibit velocity gradients along the outflow directions but do not show any clear velocity gradient perpendicular to the outflow directions.  
In the other four sources (L1527 IRS, L1448-mm, TMC-1A, and L1489 IRS) there are clear velocity gradients perpendicular to the outflow directions,  
and the magnitudes of the velocity gradients in the Class I sources (TMC-1A and L1489 IRS) are the largest,
as shown by the contrast of the color scales in Figure \ref{moment}. 
The sources with the smaller amounts of the velocity gradients perpendicular to the outflow directions (B335, IRAS 4B, L1527 IRS, and L1448-mm) all have high inclination angles ($i >$ 70\degr),  
and thus their smaller amounts of the velocity gradients are unlikely due to the projection effects. 
Our observed samples demonstrate that on an 1000 AU scale, 
Class 0 sources tend to exhibit velocity gradients along the outflow directions clearer than those perpendicular to the outflow directions, 
while Class I sources tend to exhibit dominant velocity gradients perpendicular to the outflow directions. 
Such a tendency has also been identified in the interferometric C$^{18}$O (1--0) observations on thousands of AU scale around protostars by Arce \& Sargent (2006).

\section{Analysis}
\subsection{Method to Derive Rotational Profiles}
The variety of the velocity gradients in the protostellar envelopes along and perpendicular to the outflow directions found with the present SMA observations are intriguing. 
However, 
it is in general not straightforward to attribute an observed velocity gradient to a certain systematic gas motion (i.e., rotation, infall, or outflow), 
because a velocity gradient often consists of multiple systematic motions and depends on the envelope morphologies. 
For simplicity, in the rest of this paper, we assume that the protostellar envelopes are axisymmetric around the outflow axes. 
On this assumption,
the observed velocity gradients perpendicular to the outflow directions should primarily trace the rotational motions of the envelopes around the protostars. 
In contrast, 
the velocity gradients along the outflow directions should be originated either from the outflows or the infalling motions of the envelopes, or both. 
Thus, 
for our analysis we adopt the C$^{18}$O (2--1) position--velocity (P--V) diagrams perpendicular to the outflow directions passing through the protostellar positions, produced from the combined and subcompact image cubes (Fig.~\ref{pv}).
Such diagrams can provide velocity structures of rotational motions with the least contamination from outflows, 
and infalling motions do not induce any velocity gradient in such diagrams (see Appendix B).

In the P--V diagram of B335, 
there is no clear positional offset between the blueshifted and redshifted peaks. 
In both the combined and subcompact P--V diagrams of IRAS 4B, 
a single peak at the central position around the systemic velocity is seen. 
These results show the absence of detectable rotational motions in these two sources. 
For the rest of the sources,
velocity gradients are clearly seen in their P--V diagrams. 
In these sources, 
the emission at higher velocities ($|V| >$ 0.5 -- 1 km s$^{-1}$) appears closer to the protostellar positions while the emission at offsets larger than $\sim 3\arcsec$ from the protostars shows lower velocities ($|V| <$ 0.5 -- 1 km s$^{-1}$). 
This feature implies that the rotational velocities increase as the radii decrease, i.e., spin-up rotation, 
which has been observed around other protostellar sources such as L1551 IRS 5 (Momose et al~1998; Takakuwa et al.~2004) and L1551 NE (Takakuwa et al.~2012). 

For the sources with the detectable velocity gradients perpendicular to the outflow directions, 
the rotational velocities as a function of radius are measured from the P--V diagrams. 
Figure \ref{demo} demonstrates our method. 
We adopt different fitting methods to the emission distributions in the P--V diagrams at the inner ($r < 500$ AU) and outer ($r > 500$ AU) parts. 
The borderline between the inner and outer parts ($r = 500$ AU) is chosen to be roughly the turning point of the trajectory of the spin-up rotation. 
For the inner part, 
a Gaussian function is fitted to the intensity distribution at each velocity channel in the combined P--V diagrams, to measure its peak position and hence the rotational radius at that rotational velocity (Fig.~\ref{demo} (a) \& (b)). 
For the outer part, 
a Gaussian function is fitted to the spectrum at a given position in the P--V diagrams of the subcompact data with the higher velocity resolution, to measure its centroid velocity thus the rotational velocity at that rotational radius (Fig.~\ref{demo} (c) \& (d)).
The spatial interval to perform the spectrum fitting is chosen to be the half of the beam size. 
To minimize the possible effect of the missing flux, 
the fitting is only performed at radii $\lesssim$ 10$\arcsec$
since our observations are insensitive to structures larger than 20$\arcsec$ scale at a 50\% level (Wilner \& Welch 1994).
The measured data points with the error bars are shown in the P--V diagrams of L1527 IRS, L1448-mm, TMC-1A, and L1489 IRS (Fig.~\ref{pv}). 
The rotational velocities as a function of radius after the correction of the inclination effects, as well as the results of the power-law fitting to the measured data points (i.e., $V_{\rm rot} \propto r^p$), are presented in Figure \ref{vr}. 
The best-fit power-law indices and their uncertainties are listed in Table \ref{power}.  
The estimation of the uncertainties of the data points in the P--V diagrams and the best-fit power-law indices is described in the next subsection. 
Among the four sources with the detectable velocity gradients, 
TMC-1A and L1489 IRS exhibit flatter rotational profiles on an 1000 AU scale than L1527 IRS and L1448-mm,  
and the rotational velocities in TMC-1A and L1489 IRS are higher than those in L1527 IRS and L1448-mm at the same radii. 

\subsection{Uncertainties of the Measured Rotational Profiles}
Before we discuss the variety of the rotational profiles of the protostellar envelopes and the interpretation, 
it is necessary to quantitatively assess the possible errors of the rotational profiles derived from our method. 
The sources of errors should be both the over-simplification of our method to derive rotational profiles, and the finite observational resolutions and noises. 
In this subsection, 
the uncertainty of the rotational profiles derived from our method, due to both the over-simplification of our method and the observational limitations, will be discussed.

To examine the possible effect of the over-simplification of our method, 
we constructed axisymmetric models of protostellar envelopes (e.g., Shirley et al.~2000; Harvey et al.~2003; Brinch et al.~2007b), 
and performed radiative transfer calculations of the model envelopes and SMA observing simulations of the model image cubes. 
Then, our method to derive the rotational profiles was applied to the model P--V diagrams and the resultant rotational profiles were compared to the "genuine’" rotational profiles. 
Details of these model calculations and simulations are described in Appendix B. 
First, a model of an infalling envelope without any rotational motion was tested, 
and it is confirmed that in this case our method does not deduce any "artificial" rotational profile. 
Second, a model of a rotating envelope without any infalling motion was tested, 
and it is verified that the rotational power-law index derived from our method is consistent with the "real" power-law index. 
Third, several cases of envelopes with both infalling and rotational motions were tested. 
In these cases, the synthetic vectors of the rotational and infalling motions can contribute to the line-of-sight velocities along the envelope major axes, 
and thus the P--V diagrams along the major axes do not reflect only the rotational motions but also the infalling motions. 
In these cases our method could produce artificially shallower rotational power-law indices than the genuine rotational power-law indices (i.e., $p_{\rm ob} > p_{\rm real}$). 
The magnitude of the distortion of the rotational power-law indices depends on the vertical structures ("thickness") of the envelopes, inclinations, ratios between the infalling and rotating motions, and the observational angular resolutions.  

In the case of L1527 IRS and L1448-mm there are velocity gradients along the outflow directions (see Fig.~\ref{moment}, \ref{L1527}, and \ref{L1448}), 
which could be due to the infalling motions of the envelopes. 
Thus, the "genuine" rotational power-law indices in these sources could be steeper than those listed in Table \ref{power}. 
To estimate the level of the possible distortion of the rotational profiles of L1527 IRS and L1448-mm, 
we constructed models of infalling and rotating flattened envelopes, 
and adopted the protostellar masses and the inclinations of L1527 IRS and L1448-mm reported in the literatures (Ohashi et al.~1997a; Girart \& Acord 2001; Tobin et al.~2008; Hirano et al.~2010) and the rotational profiles we measured as model parameters. 
Then we generated the model P--V diagrams and applied our method to measure the rotational power-law indices.
These simulations show that the magnitude of the possible distortion of the rotational power-law indices of L1527 IRS and L1448-mm ranges from 2$\sigma$ (i.e., $p_{\rm real} = p_{\rm ob} - 2\sigma$) to no distortion, where 1$\sigma$ is the uncertainty of the power-law indices due to the finite observational resolutions and noises.  
On the other hand there are no clear velocity gradients along the outflow directions in L1489 IRS and TMC-1A, 
and thus these sources are likely close to the "pure rotation" case. 
In conclusion, 
although the contaminations from the infalling motions could distort the rotational power-law indices of L1527 IRS and L1448-mm to artificially shallower values, 
the differences of the rotational profiles among our sources, i.e., no detectable rotation in B335 and IRAS 4B, $V_{\rm rot} \propto r^{-1}$ or steeper in L1527 IRS and L1448-mm, and $V_{\rm rot} \propto r^{-0.5}$ in L1489 IRS and TMC-1A, are still valid.

Next, we will examine the effects of the observational limitations on the rotational profiles deduced from our method. 
In our method either an intensity distribution at a given velocity channel or a spectrum at a given position in the P--V diagrams is fitted with a Gaussian function. 
In the former case the velocity channel width can be regarded as the uncertainty of the rotational velocity, 
while in the latter case the observational spatial resolution as the uncertainty of the rotational radius.
The uncertainties of the rotational radius in the former case and the rotational velocity in the later case were estimated by a Monte-Carlo method. 
Artificial random noises at the same level as that of our real observational noises were added to the observational P--V diagrams, 
and then our method was applied to the modified P--V diagrams with the additional artificial random noises. 
This process was repeated for 10,000 times, 
and the probability distributions of the rotational radius and velocity were obtained. 
The widths of the probability distributions demonstrate how sensitive the measurement of the rotational radius and velocity are to the noises, 
and provide the 1$\sigma$ uncertainties. 
To derive the errors of the power-law indices of the rotational profiles, 
we varied each data point (i.e., the rotational radius and velocity) randomly within its uncertainty, performed the power-law fitting to these varied measurements with weights by their uncertainties, and repeated this process for 10,000 times.  
Then a Gaussian function was fitted to the obtained probability distribution of the power-law index,   
and the peak position and the 1$\sigma$ width were adopted as the best-fit value and the uncertainty of the power-law index, respectively. 
We also varied the number of the repetitions by a factor of five, 
and we confirmed that 10,000 times are sufficiently large to ensure the statistical significance. 

Our analysis is based on the two-dimensional P--V diagrams,  
and an error of the adopted position angles of the outflow axes may also introduce an additional source of uncertainties to our analysis. 
To examine the uncertainty due to the error of the outflow directions, 
the position angles of the P--V diagrams were changed by 10$\degr$, 
and the same method was applied. 
We found that the variation of the best-fit power-law indices due to the 10$\degr$ error of the outflow directions is less than the uncertainty due to the random noises.   
Indeed, 
the differences between the adopted outflow directions and the minor axes of the millimeter continuum emission observed with the SMA at an angular resolution $\lesssim$ 1$\arcsec$ (Brinch et al.~2007a; Yen et al.~in prep.) are less than 10$\degr$. 
We also note that errors of the adopted $V_{\rm sys}$ could influence the fitting results. 
It is found that an error of 0.1 km s$^{-1}$ in $V_{\rm sys}$ varies the best-fit value of the power-law indices by 10\% -- 15\%, 
comparable to the uncertainty due to the random noise. 
$V_{\rm sys}$ of our target sources were measured at velocity resolutions ranging from $\sim$ 0.1 to $\sim$ 0.3 km s$^{-1}$ (Ohashi et al.~1997a; Hogerheijde et al.~1998; Curiel et al.~1999; Yen et al.~2011), except for that of L1448-mm, 
which was measured at a velocity resolution of $\sim$ 0.6 km s$^{-1}$ (Volgenau et al.~2006).  
The accuracy of the measured $V_{\rm sys}$ can be expressed as $\sim$ ${\Delta V}/(S/N)$, where $\Delta V$ denotes the velocity resolution and $S/N$ the signal-to-noise ratio of the line emission, and is found to be better than 0.1 km s$^{-1}$ for all of our sources. 
In L1527 IRS, 
the adopted $V_{\rm sys}$ ($= 5.7$ km s$^{-1}$) is inbetween the central two peaks at $V \sim -0.3$ and 0.2 km s$^{-1}$ in the subcompact P--V diagram. 
In L1448-mm, 
$V_{\rm sys}$ is coincident with the velocity of the emission peak in the P--V diagram. 
In TMC-1A and L1489 IRS, 
the mean velocities of the C$^{18}$O (2--1) emission at the protostellar positions are consistent with $V_{\rm sys}$ (Fig.~\ref{moment}). 
These results suggest that $V_{\rm sys}$ measured by the single-dish observations match those of the inner regions observed with the SMA, 
and that the errors of the adopted $V_{\rm sys}$ are unlikely to affect the estimates of the rotational power-law indices.

Finally, we note the present method is based on the assumption that the protostellar envelopes are axisymmetric around their rotational axes. If protostellar envelopes are non-axisymmetric and filamentary for example, 
the velocity gradients in the P--V diagrams perpendicular to the outflow directions could be caused by other kinematics instead of rotational motions, such as infalling flow along the filaments (Tobin et al.~2012b).

\section{Discussion}
\subsection{Variations of the Rotational Motions around the Protostellar Sources}
Our SMA observations of six representative protostellar sources and the data analysis have unveiled the variations of the rotational motions of the envelopes on the 100 to 1000 AU scales. 
Toward B335 and IRAS 4B no detectable velocity gradients perpendicular to the outflow directions are seen, 
suggesting absence of the rotational motions in the envelopes. 
It is possible that in these sources CO depletion occurs in the envelope mid-planes and thus the rotational motions are not detected with the present C$^{18}$O (2--1) observations. 
However, in B335 the C$^{18}$O (2--1) emission is elongated perpendicularly to the outflow direction around the systemic velocity, 
similar to the 1.3 mm dust continuum emission (Yen et al.~2010; 2011). 
Furthermore, the temperature of the envelope mid-plane in B335 is estimated to be higher than 20 K, the CO sublimation temperature (Yen et al.~2011), 
and the envelope around IRAS 4B has been identified as a "hot corino" ($T_{\rm k}$ $\sim$ 100 K; Bottinelli et al.~2007). 
Therefore, 
it is unlikely that CO depletion occurs and affects the observed kinematics in the envelopes around B335 and IRAS 4B. 
Although there is possible distortion of the rotational power-law indices of L1527 IRS and L1448-mm due to the contamination from the infalling motions as discussed above, 
the difference of the rotational profiles from those around the other sources is likely real. 
In the case of L1489 IRS and TMC-1A no infalling motion is identified in our data and thus the derived rotational power-law indices are probably valid. 
In this subsection, the variations of the rotational motions in the protostellar envelopes and their possible interpretation will be discussed.

Figure \ref{vrdata} plots of our measured rotational velocities and radii. 
The envelopes around B335 and IRAS 4B do not show any sign of rotational motion in our SMA data, 
and the detection limit of the rotational velocities is drawn as a red dotted line in Figure \ref{vrdata} (see Yen et al.~2010 for details of the detection limit). 
Our previous SMA observations of B335 in the C$^{18}$O (2--1) emission have found infalling motion in the 1000 AU scale envelope (Yen et al.~2010; 2011). 
In IRAS 4B, the inverse P Cygni profile seen in the H$_2$CO (3$_{12}$--2$_{11}$) emission suggests the presence of infalling motion (Di Francesco et al.~2001).
Therefore, 
in the envelopes around B335 and IRAS 4B there are infalling motions but little rotation. 
The envelopes around the other sources (L1527 IRS, L1448-mm, TMC-1A, and L1489 IRS) show detectable rotational motions. 
The rotational velocities around L1489 IRS and TMC-1A are higher than those around L1527 IRS and L1448-mm at the same radii, and the rotational profiles of L1489 IRS and TMC-1A ($p \sim -0.5$) are flatter than those of L1527 IRS and L1448-mm ($p \sim -1$ or steeper). 
A rotational profile with $p = -1$ denotes rotational motion with a conserved angular momentum in a dynamically infalling envelope (Ulrich 1976; Cassen \& Moosman 1981; Terebey et al.~1984; Basu et al.~1998). 
In fact, previous interferometric molecular-line observations of L1527 IRS (Ohashi et al.~1997a) and L1448-mm (Curiel et al.~1999) have identified the infalling motions in the envelopes. 
On the other hand, a rotational profile with $p = -0.5$ implies Keplerian rotation, 
and the presence of a Keplerian disk in L1489 IRS has been reported by Brinch et al.~(2007a).

For comparison, rotational velocities around the other Class 0 and I protostars taken from the literatures are also plotted in Figure \ref{vrdata}. 
Around Class I protostars IRS 63, Elias 29, and L1551 NE, 
the presence of Keplerian disks has been reported with the SMA observations (Lommen et al.~2008; Takakuwa et al.~2012), 
and the rotational velocities at the disk outer radii appear to be consistent with those around L1489 IRS and TMC-1A. 
Around a Class I protostar HH 111, Lee (2010) measured the rotational velocities of the surrounding envelope and claimed the presence of the transition of the rotational profile from $V_{\rm rot} \propto r^{-1}$ to $V_{\rm rot} \propto r^{-0.5}$ (black solid lines in Figure \ref{vrdata}). 
The reported Keplerian rotation at inner radii around HH 111 also appears to be consistent with the Keplerian rotation around L1489 IRS and TMC-1A. 
On the other hand, 
the presence of a possible Keplerian disk around L1527 IRS has been reported (Tobin et al.~2012a), 
and the rotational velocity at the disk outer radius is consistent with that from the extrapolation of our derived rotational profile of L1527 IRS to the inner radii. 
Infalling motions in the surrounding envelopes have been identified around Class I protostar L1551 IRS 5 (Momose et al.~1998) and Class 0 protostar HH 212 (Lee et al.~2006), 
and their rotational velocities are consistent with those around L1448-mm and L1527 IRS. 
Around Class 0 protostars HH 211 and L1157-mm, rigid-body rotation on thousands of AU scale in the protostellar envelopes have been proposed (Chiang et al. 2010; Tanner \& Arce 2011), 
while their rotational velocities at radii of few thousand AU appear to be consistent with those around L1448-mm and L1527 IRS. 
Thus, the rotational velocities of the Keplerian disks around the Class I sources are higher than those around the Class 0 and I sources exhibiting non-Keplerian rotation on hundreds of AU scales. 
Our measured rotational profiles combined with the rotational velocities reported in the literatures  
demonstrate that the rotational motions of the envelopes on the 100 to 1000 AU scales around the protostars can be classified into the three categories; 1) no detectable rotation but infalling motions (Class 0 protostars, B335 and IRAS 4B), 2) rotational motions with $V_{\rm rot} \propto r^{-1}$ in the infalling envelopes (Class 0 and I protostars, L1527 IRS, L1448-mm, L1551 IRS 5, and HH 212), 3) Keplerian rotation (Class I protostars, L1489 IRS, TMC-1A, IRS 63, Elias 29, HH 111, and L1551 NE), 
and the protostellar sources in the later evolutionary stages exhibit faster rotational motions.  

In the following, we will show that the three categories of the rotational motions of the envelopes can be interpreted with a conventional, analytical picture of evolution of a protostellar envelope into Keplerian-disk formation (e.g., Shu et al.~1987).
In the conventional picture of an infalling envelope, 
each infalling material conserves its angular momentum, 
and its rotational velocity ($v_\phi$) is described as, 
\begin{equation} 
v_\phi = \sqrt{(\frac{GM}{r})(1-\frac{\cos\theta}{\cos\theta_0})}\frac{\sin\theta_0}{\sin\theta}, \label{vrot1}
\end{equation}
and 
\begin{equation} 
r = \frac{r_{\rm d}\cos\theta_0\sin^2\theta_0}{\cos\theta_0-\cos\theta} = \frac{r_{\rm d}\sin^2\theta_0}{1-\cos\theta/\cos\theta_0}, \label{vrot2}
\end{equation}
where $G$ is the gravitational constant, $M$ is the central protostellar mass, $r_{\rm d}$ is the centrifugal radius in mid-plane, and $\theta_0$ is the initial polar angle of the infalling material (see Equation 9 \& 10 in Ulrich 1976). 
Therefore, the rotational velocity can be expressed as $v_\phi = \sqrt{GMr_{\rm d}}(\sin^2\theta_0/\sin\theta)r^{-1}$, which yields a rotational profile with a power-law index of $-1$ close to the mid-plane (i.e., $\sin^2\theta_0 / \sin\theta \sim 1$).
The rotational velocity of the infalling material increases as the material approaches to the central protostar, 
and eventually becomes comparable to the infalling velocity. 
At this point, 
the infalling material becomes centrifugally supported and forms a Keplerian disk. 
The presence of such a rotational profile, $p = -0.5$ at the inner radii and $p = -1$ at the outer radii, has been reported in HH 111 (Lee 2010).
When the inside-out collapse takes place in an initial dense core with rigid-body rotation as seen in the NH$_3$ cores (Goodman et al.~1993), 
the innermost part of the dense core with the smallest angular momentum collapses first. 
Because of the smallest angular momentum, 
the rotational velocity of the infalling material is small even though it becomes larger in the vicinity of the central protostar. 
As a result, most of the infalling materials accrete onto the central protostar without forming a Keplerian disk. 
As the expansion wave propagates outward in the dense core, 
an outer part of the dense core with a larger angular momentum starts to collapse. 
Because of the larger angular momentum, 
the rotational velocity becomes large enough to form a Keplerian disk before the infalling material accretes onto the central protostar.
When the expansion wave propagates further, materials with even larger angular momenta start to collapse, 
and they become centrifugally supported at larger radii, expanding the size of the Keplerian disk. 

To present this scenario analytically, 
we followed the calculation by Terebey et al.~(1984) and Basu et al.~(1998), and calculated the expected rotational profiles at different evolutionary stages of the collapse of a dense core. 
The initial condition is assumed to be an isothermal singular sphere with a rigid-body rotation. 
The initial density profile of the sphere $\rho(r)$ is
\begin{equation}
\rho(r) = \frac{{c_{\rm s}}^2}{2\pi G}r^{-2}, 
\end{equation}
where $c_{\rm s}$ is the sound speed (Shu 1977).  
The enclosed mass as a function of radius $M(r)$ is given by
\begin{equation}
M(r) = \frac{2{c_{\rm s}}^2}{G}r.
\end{equation} 
The initial distribution of the specific angular momenta $j(r)$ is 
\begin{equation}
j(r) = \omega r^2, \label{jr}
\end{equation}
where $\omega$ is the angular velocity of the rigid-body rotation. 
On the assumption that the angular momentum is conserved during the collapse, 
the Keplerian radius ($r_{\rm d}$) of each mass shell initially at a radius of $r$ can be expressed as, 
\begin{equation}
\frac{{j(r)}^2}{{r_{\rm d}}^3} = \frac{GM(r)}{{r_{\rm d}}^2}. \label{rd}
\end{equation}
On the assumption of the steady mass accretion ($\equiv \dot{M}$), 
the Keplerian radius as a function of time $t$ can be derived from Equation \ref{rd} as
\begin{equation}
r_{\rm d} = \frac{{j(r)}^2}{GM(r)} = \frac{\omega^2 G^3 {M(r)}^3}{16{c_{\rm s}}^8} = \frac{\omega^2 G^3 {\dot{M}}^3}{16{c_{\rm s}}^8}t^{3}. \label{rdt}  
\end{equation}
This is the solution by Terebey et al.~(1984). 
Basu et al.~(1998) performed the same calculation with the initial rotation of a constant velocity, 
and their solution is $r_{\rm d} \propto t$. 
Equation \ref{rdt} implies that when a central protostar + disk system increases its mass to $M(r)$ after a time interval of $t = M(r)/\dot{M}$, 
the material initially at a radius of $r$ has collapsed and carried an angular momentum of $j(r)$ toward the center,   
and the rotational velocity of the material reaches its Keplerian velocity at a radius of $r_{\rm d}$.  
On the assumption that the gravity is dominated by the central protostar, 
the rotational profile follows $V_{\rm rot} \propto r^{-0.5}$ inside $r_{\rm d}$, i.e., Keplerian rotation. 
In the dynamically infalling region and outside the Keplerian radius (i.e., $r_{\rm d} < r < c_{\rm s}t$), 
the infalling material is accelerated to supersonic, 
and the initial profile of the specific angular momentum shown by Equation \ref{jr} is altered.
Then, the radial profile of the specific angular momentum becomes constant asymptotically toward the inner radii (see Figure 2 in Terebey et al.~1984),  
and the rotational profile of the infalling material becomes $V_{\rm rot} \propto r^{-1}$ as derived from Equation \ref{vrot1} and \ref{vrot2}.
Therefore, 
the rotational profile as a function of time can be described as
\begin{equation}
V_{\rm rot}(r) = \sqrt{\frac{G\dot{M}t}{{r_{\rm d}}^2}} \times
\begin{cases}
( \frac{r}{r_{\rm d}})^{-0.5} & r \leq r_{\rm d} \\
(\frac{r}{r_{\rm d}})^{-1.0} & r_{\rm d} \leq r < r_{\rm infall} 
\end{cases}
,  \label{vrt}
\end{equation}
where $r_{\rm d}$ is evaluated by Equation \ref{rdt} and $r_{\rm infall}$ is the location of the expansion wave ($= c_{\rm s}t$; Shu 1977). 
Note that the material located closer to the polar axis possesses a smaller angular momentum, 
and the calculated rotational profile is applied to the material on the equatorial plane. 
Also note that the effect of the magnetic field and outflows, and the mass distribution inside $r_{\rm d}$ are not considered in this calculation. 

The expected rotational profiles at different collapse stages calculated from Equation \ref{vrt} are shown as solid lines in Figure \ref{rdisk}. 
Here the sound speed $c_{\rm s}$ is assumed to be $\sim$ 0.2 km s$^{-1}$ (i.e., the gas kinematic temperature is 10 K), 
and thus the mass accretion rate $\dot{M}$ (= $c_{\rm s}^3/G$) is $\sim$ 2 $\times$ 10$^{-6}$ $M_{\sun}$ yr$^{-1}$ (Shu 1977).  
The angular velocity $\omega$ is assumed to be 6.5 $\times$ 10$^{-14}$ s$^{-1}$, 
which corresponds to 2 km s$^{-1}$ pc$^{-1}$ and is similar to the mean magnitude of the large-scale rotational motions of dense cores and protostellar envelopes within a factor of two (Goodman et al.~1993; Gaselli et al.~2002; Tobin et al.~2011).
Note that the expected rotational profile is sensitive to the initial conditions,   
and these parameters are adopted with an intent to demonstrate the evolutionary trend rather than to constrain the initial conditions of star formation. 
As long as the initial rotational profile shows an increasing specific angular momentum with the increasing radius, 
the overall evolutionary trend deduced here is valid, 
although the time dependence should be different. 
In the calculation the steady mass accretion is assumed, 
although the mass accretion is likely episodic and its rate may change with time (e.g., Evans et al.~2009).  
A time-dependent mass accretion rate do not affect the evolutionary trend of the rotational profile described by Equation \ref{vrt}. 

As shown in Figure \ref{rdisk}, 
the rotational velocities of our model increase with time. 
At $t < 5 \times 10^{4}$ yr, the rotational velocities are slow and below our detection limit.  
This applies to the case of B335 and IRAS 4B, 
where their rotational motions are not detected but the infalling motions are likely present.  
At $t < 1.5 \times 10^{5}$ yr, 
the radius of the Keplerian disk is less than 45 AU and cannot be seen in this plot,  
and the material at radii between 45 and 6000 AU exhibits a rotational profile with a power-law index of $-1$. 
L1527 IRS and L1448-mm, which exhibit steep rotational profiles on the 100 to 1000 AU scales, are fit to this stage. 
Recent observations of L1527 IRS in the C$^{18}$O (2--1) emission line with the Atacama Large Millimeter/Submillimeter Array (ALMA) have revealed that the rotational profile with a power-law index of $-1$ extends down to $\sim$ 50 AU (Ohashi et al.~2013 in prep.), consistent with the expected rotational profile at $t \sim 1.5 \times 10^{5}$ years. 
On the other hand, 
recent CARMA observations of L1527 IRS have found a possible Keplerian disk with a radius of $\sim90$ AU (Tobin et al. 2012a). 
Although the reason for the possible discrepancy between the ALMA and CARMA results is unclear, 
the presence of the Keplerian disk with the radius of $\sim90$ as well as our measured rotational profile with the SMA can be explained by our model with a factor of two to three larger $\omega$.
At $t = 2.5 \times 10^{5}$ yr, 
the radius of the Keplerian disk, shown as a breaking point in the rotational profile, grows to $\sim$ 200 AU.  
The power-law index of the rotational profile at radii less than 200 AU has changed from $-1$ to $-0.5$, 
and that at radii larger than 200 AU remains $-1$. 
At $t = 5 \times 10^{5}$ yr, 
the radius of the Keplerian disk grows to more than 1000 AU, 
which is consistent with the data points of L1489 IRS and TMC-1A. 
Thus, the evolutionary sequence of the rotational profile predicted by our simple calculations matches the observed rotational profiles.

We stress here that the time scales shown in our analytical calculation are merely an indicator of the collapse stages, 
and do not represent the real ages of the individual sources. 
As discussed above, the evolutionary time scale is sensitive to the initial conditions of dense cores, 
such as the initial angular velocity, the initial rotational profile, and the sound speed and hence the mass accretion rate. 
In fact, among our sample L1448-mm is a factor of two to five more luminous than the other sources and drives energetic molecular outflows (Bachiller 1990; Hirano et al.~2010), 
suggesting that the mass accretion rate onto L1448-mm is higher than that onto the other sources. 
Furthermore, the locations of our target sources span different types of star-forming regions from an isolated Bok globule, an isolated star-forming region in Taurus, to a cluster-forming region in Perseus. 
The different environments are likely to affect the initial conditions of the dense cores and hence the real time scale of the evolution of the protostellar envelopes. 
On the other hand, 
we have unveiled three categories of the envelope kinematics from our target sources as well as the sources from the literatures as described above. 
It is unlikely that these categories merely reflect the "individuality" of each source. 
We suggest that the three types of the envelope kinematics reflect evolution of the protostellar envelopes, 
and that our simple analytical model provides at least a qualitative guideline of evolution of protostellar envelopes to Keplerian-disk formation.

{\subsection{Observations and MHD Simulations of Disk Formation}}
Recent MHD simulations of disk formation in a collapsing dense core have shown that 
if the mass-to-flux ratio is less than 10 times of the critical value ($1/2\pi \sqrt{G}$; Nakano \& Nakamura 1978), 
the magnetic field can efficiently remove the angular momentum of the collapsing material and suppress the growth of a centrifugally-supported disk  (e.g., Mellon \& Li 2008, 2009; Machida et al.~2011; Li et al.~2011; Joos et al.~2012; Dapp et al.~2012). 
Because a part of the angular momentum of the collapsing material is removed by magnetic braking, 
these MHD simulations predict a shallower ($p > -1$) rotational profile of infalling envelopes, 
and its power-law index depends on the efficiency of magnetic braking (e.g., Mellon \& Li 2008, 2009). 
Observations of the Zeeman splitting of the OH line toward 34 dark cloud cores show that the mean mass-to-flux ratio is $\sim$ 4 -- 5 (Troland \& Crutcher 2008), 
which probes the outer parts of the cores with a mean gas density of $10^{3}-10^{4}$ cm$^{-3}$. 
Observations of the Zeeman splitting of the CN line toward massive star-forming regions have shown that the mean mass-to-flux ratio is 1-- 4 in the denser regions ($n_{\rm H_2} \sim 10^4 -10^6$ cm$^{-3}$), 
but the result may not be applied to the case of low-mass star-forming regions where the physical conditions are likely different from those in the massive star-forming regions.
In the MHD simulations adopting these observed mass-to-flux ratios, 
efficient magnetic braking restricts the radii of Keplerian disks to be less than 10 AU.  
Incorporation of non-ideal MHD effects in the simulations cannot reduce the efficiency of magnetic braking to the level that Keplerian disks with radii of hundreds of AU can form (e.g., Mellon \& Li 2009; Li et al.~2011; Dapp et al.~2012). 
The dissipation of protostellar envelopes due to outflows or mass accretion, which reduces the efficiency of magnetic braking, is proposed to enable the formation of large-scale Keplerian disks (e.g., Mellon \& Li 2008; Machida et al.~2011). 
Machida et al. (2011) have calculated the evolution of magnetized collapsing cores until the stage when most of the mass of the envelopes are ejected or accreted, 
and have formed centrifugally-supported disks with radii more than 100 AU. 
However, 
in their simulations, 
the centrifugally-supported disks are more massive than the protostars by a factor of two to five. 
Such massive disks have not been observed around Class I protostars, 
and their rotational motions deviate from Keplerian rotation and show $p > -0.5$ due to the self-gravity. 
The other possible mechanism to reduce the efficiency of magnetic braking is the misalignment between the direction of the magnetic field and the rotational axes of dense cores (e.g., Joos et al.~2012). 
Recent CARMA survey of polarized dust continuum emission at 1.3 mm shows that the inferred directions of the magnetic field in low-mass protostellar sources are not tightly aligned with the outflow directions on an 1000 AU scale (Hull et al.~2012), 
which could support this scenario to reduce the efficiency of magnetic braking.

Our SMA observations of L1448-mm, L1527 IRS, TMC-1A, and L1489 IRS found that the angular momenta of the collapsing cores are conserved and efficiently transferred to the inner radii, 
and the centrifugally-supported disks with radii of hundreds of AU have been formed in the Class I stage. 
These results suggest that magnetic braking has already become minimal in the Class 0 to Class I stages. 
On the other hand, 
the absence of detectable rotational motions in B335 and IRAS 4B, which are in the early collapse stage, could suggest the presence of magnetic braking in the early collapse stage. 
The efficiency of magnetic braking and its time evolution in protostellar envelopes are still controversial,
and currently there is no direct measurement of the mass-to-flux ratio of protostellar envelopes on an 1000 AU scale around low-mass protostars. 
The magnetic fields toward low-mass protostellar sources are mostly measured with single-dish observations at an angular resolution of $\sim$ 10$\arcsec$ ($>$ 1400 AU; e.g., Attard et al.~2009; Davidson et al.~2011), insufficient to resolve the structures of the magnetic fields on the 1000 AU scale. 
The magnetic field in NGC 1333 IRAS 4A, a low-mass protobinary system with a separation of $\sim$ 2$\arcsec$ (500 AU), has been observed at a high angular resolution of 1\farcs6 (400 AU; Girart et al.~2006).  
However, NGC 1333 IRAS 4A  drives two outflows in different directions (e.g., Choi 2005),  
and hence the relation between the kinematics of the inner envelope and the magnetic field is likely complicated. 
To investigate the role of the magnetic field in the formation of Keplerian disks,   
observations revealing the structure and strength of the magnetic field on hundreds of AU scale around low-mass protostars as well as precise measurements of envelope rotational profiles for a statistically significant sample are crucial. 
Recently Koch et al.~(2012a, b) developed a method to measure the mass-to-flux ratio by the gradients of polarization directions and intensity distributions by using the SMA data of high-mass star-forming regions.  
Such studies can be applied to inner envelopes around low-mass protostars with future ALMA observations. 

\subsection{Keplerian Disks around Class I Protostars} 
Our measured rotational power-law indices suggest that TMC-1A and L1489 IRS are surrounded by Keplerian disks. 
From the Keplerian rotational profile ($V_{\rm rot} = \sqrt{GM_{*}}r^{-0.5}$), 
the central protostellar masses are estimated to be $1.1\pm0.1\ M_\sun$ in TMC-1A and $1.8\pm0.2\ M_\sun$ in L1489 IRS on the assumption that the inclination angles of the Keplerian disks are 30$\degr$ and 50$\degr$, respectively (Table \ref{sample}).  
From the total flux of the 1.3 mm continuum emission, 
the masses of the Keplerian disks around TMC-1A and L1489 IRS are estimated to be 0.06 -- 0.025 $M_\sun$ and 0.02 -- 0.007 $M_\sun$ with dust temperatures of 15 -- 30 K (Appendix A). 
The estimated masses are approximately consistent with previous estimates of the disk masses in TMC-1A (0.035 -- 0.005 $M_\sun$; J{\o}rgensen et al.~2009; Eisner 2012) and L1489 IRS (0.018 -- 0.004 $M_\sun$; Brinch et al.~2007a; J{\o}rgensen et al.~2009; Eisner 2012).
The disk masses around TMC-1A and L1489 IRS are orders of magnitude lower than the protostellar masses, 
and hence the disk self-gravity is indeed negligible. 
Keplerian disks around more evolved sources (i.e., T Tauri stars) have masses ranging from 10$^{-1}$ to 10$^{-4}$ $M_\sun$ (e.g., Guilloteau et al.~2011) and radii ranging from 100 to 800 AU (e.g., Simon et al.~2000). 
Our results of TMC-1A and L1489 IRS suggest that Keplerian disks with sizes and masses comparable to those of classical disks have formed in the Class I stage. 
The presence of similar Keplerian disks around other Class I protostars has also been reported (Brinch et al.~2007a; Lommen et al.~2008; J{\o}rgensen et al.~2009; Takakuwa et al.~2012). 

If the mass accretion rate is constant ($\sim c_{\rm s}^3/G$; Shu 1977) in TMC-1A and L1489 IRS,   
their ages ($\equiv M_*/\dot{M}$) are estimated to be 6 $\times$ 10$^{5}$ and 9 $\times$ 10$^{5}$ yr, respectively.
In the MHD simulations made by Machida et al.~(2011), 
at $\sim$ 10$^{5}$ yr, 
the radii of centrifugally-supported disks increase to 100 -- 1000 AU (depending on the mass-to-flux ratio), 
which are comparable to the radii of the Keplerian disks in TMC-1A and L1489 IRS. 
However,     
the masses of the centrifugally-supported disks in the MHD simulations grow to $\gtrsim$ 0.1 $M_\sun$ and are larger than the protostellar mass by a factor of two to five.  
Further theoretical efforts are required to reproduce the observationally-identified Keplerian disks around Class I protostars. 

\section{Summary}
We have performed imaging and analysis of the SMA observations in the C$^{18}$O (2--1) emission line toward three Class 0 (B335, IRAS 4B, and L1448-mm), one Class 0/I (L1527 IRS), and two Class I protostars (TMC-1A, and L1489 IRS). 
The aim of the present research is to study the rotational motions of the protostellar envelopes on 100 to 1000 AU scales, 
and to investigate formation process of Keplerian disks.  
Our main results are summarized below.
\begin{enumerate}
\item{
The C$^{18}$O (2--1) emission shows $\sim$ 1500 -- 2000 AU scale, centrally-peaked condensations associated with the protostars plus extensions toward the outflow directions, 
and likely traces the protostellar envelopes with the contamination from the outflows. 
The protostellar envelopes around the two Class 0 protostars, B335 and IRAS 4B, show no detectable velocity gradient perpendicular to the outflow directions.
The protostellar envelopes around L1527 IRS (Class 0/I) and L1448-mm (Class 0) show substantial velocity gradients, 
and those around the Class I protostars, TMC-1A and L1489 IRS, show the largest velocity gradients among the sample.
The velocity gradients perpendicular to the outflow directions most likely trace the rotational motions of the protostellar envelopes around the protostars.}
\item{
We developed a simple analytical method to measure the rotational velocities of the protostellar envelopes as a function of radius ($V_{\rm rot} \propto r^p$) over the 100 -- 1000 AU scales from the observed P--V diagrams perpendicular to the outflow directions. 
Possible systematic errors of the measured rotational profiles due to the over-simplification of our method are evaluated with models of infalling and rotating envelopes and the observing simulations of the model images. 
With a Monte-Carlo method errors of the derived rotational profiles due to the observational noises and resolutions are also estimated. 
We found that L1527 IRS and L1448-mm exhibit rotational profiles with the power-law indices $p = -1.0 \pm 0.2$ and $-1.0 \pm 0.1$, respectively, or steeper if the contamination from the infalling motions affects the P--V diagrams perpendicular to the outflow directions. 
TMC-1A and L1489 IRS exhibit flatter rotational profiles with the power-law indices $p = -0.6 \pm 0.1$ and $-0.5 \pm 0.1$, respectively. 
B335 and IRAS 4B do not show any detectable rotational motion. 
These results, as well as previous observational results of rotational motions of other protostellar envelopes, demonstrate that the rotational motions of the envelopes on the 100 to 1000 AU scales can be classified into the three categories; 1) no detectable rotation but infalling motions (Class 0 protostars, B335 and IRAS 4B), 2) rotational motions with $V_{\rm rot} \propto r^{-1}$ in infalling envelopes (Class 0 and I protostars, L1527, L1448-mm, L1551 IRS 5, and HH 212), and 3) Keplerian rotation (Class I protostars, L1489 IRS, TMC-1A, IRS 63, Elias 29, HH 111, and L1551 NE).
}
\item{
We demonstrated that the three categories of the rotational motions of the protostellar envelopes can be reproduced with 
analytical calculations of the inside-out collapse model combined with rigid-body rotation and a conserved angular momentum. 
At the initial onset of the collapse the innermost part of the dense core with smaller angular momenta collapses first. 
As the expansion wave propagates outward, the envelope materials with larger angular momenta start to collapse. 
At the late stage of the collapse, more angular momenta transfer to the center, 
and Keplerian disks around protostars are formed. 
In reality the individual evolutions of the protostellar envelopes depend on the environmental conditions such as the magnitude of the initial rotational motion and sound speed, 
and it is difficult to sort the various protostellar sources along a common evolutionary track of a protostellar envelope. 
On the other hand, the three categories of the envelope kinematics found from our observations as well as from the literatures are unlikely to be explained merely by the "individuality" of each source. 
We suggest that the three types of the envelope kinematics reflect evolution of the protostellar envelopes, 
and that our simple analytical model provides at least a qualitative guideline of evolution of protostellar envelopes into Keplerian-disk formation.
}
\item{
Around TMC-1A and L1489 IRS Keplerian disks with the radii of hundreds of AU, comparable to those around T-Tauri stars, are present. 
The protostellar masses of TMC-1A and L1489 IRS are estimated to be $1.1 \pm 0.1$ and $1.8 \pm 0.2$ $M_\sun$, respectively, 
and the disk masses are estimated to be 0.005 -- 0.035 $M_\sun$ and 0.007 -- 0.02 $M_\sun$. 
These results as well as results from recent millimeter and submillimeter interferometric observations of other Class I sources imply that Keplerian disks are already well developed in the Class I stage. 
Theoretical studies of disk formation including the realistic magnetic fields have trouble reproducing these observationally-identified disks around Class I protostars,
because magnetic breaking efficiently suppresses the disk formation. 
Sensitive observational measurements of strength and structures of the magnetic field are essential to refine theoretical studies of disk formation and to explain an increasing number of Keplerian disks around Class I protostars.
}
\end{enumerate}

\acknowledgments
We thank the anonymous referee for insightful comments and suggestions, 
and all the SMA staffs supporting this work. 
S.T. acknowledges a grant from the National Science Council of Taiwan (NSC 99-2112-M-001-013-MY3) in support of this work. 
The research of N. O. was partially supported by NSC 99-2112-M-001-008-MY3. 

\appendix

\section{1.3 mm Continuum Images}
Figure \ref{continuum} shows the 1.3 mm continuum images of our sample sources. 
The continuum images of IRAS 4B, L1527 IRS, L1448-mm, TMC-1A, and L1489 IRS were taken with the SMA compact and subcompact configurations, 
and that of B335 was taken with the SMA compact configuration.
The spatial resolutions and the noise levels of these images are summarized in Table \ref{contable}. 
1.3 mm dust continuum emission is detected toward all the sources. 
In IRAS 4B, two dusty components, one associated with the primary source of the binary to the west (4B) and the other with the secondary (4B$^\prime$), are seen. 
In L1448-mm, in addition to the central compact component associated with the protostellar source there are a secondary peak to the southeast and an extended emission component elongated along the outflow direction. 
The secondary peak is coincident with the redshifted C$^{18}$O (2--1) clump to the southeast seen in the velocity channel maps (Fig.~\ref{L1448}).

Two-dimensional Gaussian fittings were performed to the central components associated with the protostellar sources, 
and the total fluxes, deconvolved sizes, and the position angles are derived (Table \ref{contable}). 
In IRAS 4B, B335, and L1489 IRS, the central components are elongated perpendicular to the outflow directions. 
In L1527 IRS, L1448-mm, and TMC-1A, the aspect ratios of the central components are larger than 0.75, and hence the elongation of the central components is not clear.

Toward TMC-1A and L1489 IRS Keplerian disks have been identified from the C$^{18}$O (2--1) data. 
Assuming that the 1.3 mm dust continuum emission in TMC-1A and L1489 IRS arise from the Keplerian disks, 
the masses of the Keplerian disks can be estimated from the total 1.3 mm fluxes as; 
\begin{equation}
M_{\rm d} = \frac{F_{\nu}d^{2}} {\kappa_{\rm 1.3 mm} B(T_{\rm dust})},
\end{equation}
where $F_{\nu}$ is the total flux, $\kappa_{\rm 1.3 mm}$ is the dust mass opacity at 1.3 mm, $T_{\rm dust}$ is the dust temperature, and $B$ is the Planck function. 
On the assumption that the frequency ($\nu$) dependence of the dust mass opacity ($\equiv \kappa_{\nu}$) is $\kappa_{\nu} = 0.1 \times (\nu/10^{12})^{\beta}$ (Beckwith et al.~1990), the mass opacity at 1.3 mm is 0.023 cm$^{2}$ g$^{-1}$ with $\beta = 1.0$ (e.g., J{\o}rgensen et al.~2007). 
The masses of the Keplerian disks in TMC-1A and L1489 IRS are estimated to be 0.025 -- 0.06 $M_\sun$ and 0.007 -- 0.02 $M_\sun$ with $T_{\rm dust}$ = 15 -- 30 K, which are typical temperatures of circumstellar disks on hundreds of AU scale around T Tauri stars (e.g., Isella et al.~2009; Guilloteau et al.~2011).

\section{Measuring Rotational Profiles from Position--Velocity Diagrams}
To investigate the feasibility and limitation of our method to derive rotational profiles from the observed P--V diagrams perpendicular to the outflow directions, 
model images of infalling and rotating protostellar envelopes in the C$^{18}$O (2--1) emission were constructed, 
and the SMA observing simulations with the model images were performed. 
Then, our method to derive rotational profiles was applied to the simulated P--V diagrams, 
and the outputs were compared to the "genuine" rotational profiles of the model envelopes. 
In the model envelopes the density and temperature profiles are assumed to be $\rho(r) \propto r^{-1.5} \cdot \sin^f\theta$ and $T(r) = 10 \times (r / 5000\ {\rm AU})^{-0.4}$, respectively (Shirley et al.~2000; Harvey et al.~2003; Brinch et al.~2007b). 
Here,  $\sin^f\theta$ represents the "flatness" of the model envelopes, and $f = 4, 8$, and infinity (geometrically-thin disk) are adopted. 
In the geometrically-thin case a two-dimensional Gaussian intensity distribution is adopted as the moment 0 map.  
The distance to the model envelopes is adopted to be 140 pc, same as that to our targets in the Taurus region. 
The outer radius of the model envelopes is adopted to be 4200 AU ($= 30\arcsec$), similar to the radius of the SMA primary beam. 
For simplicity, the infalling motion is assumed to follow radial free fall, $v_{\rm in} = \sqrt{2GM_* / r}$, 
while the rotational motion is assumed to have a power-law profile, $v_{\rm rot} = v_0 \cdot (R_{\rm rot} / 100\ {\rm AU})^p$, 
where $G$ is the gravitational constant, $M_*$ is the protostellar mass, and $R_{\rm rot}$ is the rotational radius (i.e., the distance to the rotational axis). 
The model image cubes in the C$^{18}$O (2--1) emission were generated with a three-dimensional radiative transfer code, LIME (Brinch \& Hogerheijde 2010), 
and those of the geometrically-thin envelopes were generated with MIRIAD tasks, $velmodel$ and $velimage$. 
The generated image cubes were sampled with the same $uv$ points and velocity resolution as those of our real observations with a MIRIAD task $uvmodel$, to produce the simulated visibility data. 
Then the simulated images were created from the simulated visibility data, 
and our method to derive rotational profiles was applied to the simulated P--V diagrams,  
which includes random noises to match the signal-to-noise ratios of the simulated P--V diagrams with those of the real observations.

A series of simulations with $M_*$ = 0, 0.05, 0.1, and 0.2 $M_\sun$, $v_0$ = 0, 1, and 4 km s$^{-1}$, $p = -0.5, -0.7, -1.0$, and $-1.3$, an inclination angle of 45$\degr$, 60$\degr$, and 90$\degr$, and $f = 4, 8$, and infinity, were performed. 
The parameter space was chosen to approximately cover the anticipated ranges of those parameters of our target sources. 
Here we show three representative cases with $f$ = 4 and an inclination angle of 60$\degr$; 
1) no rotation but infalling motion ($M_*$ = 0.2 $M_\sun$ and $v_0$ = 0); 
2) rotation without infalling motion ($M_*$ = 0, $v_0$ = 1 km s$^{-1}$, and $p = -1$); 
and 3) rotating and infalling envelope ($M_*$ = 0.2 $M_\sun$, $v_0$ = 1 km s$^{-1}$, and $p = -1$). 
Figure \ref{case} shows the model moment 0 and 1 maps, P--V diagrams, and the data points of the rotational velocities deduced from our method on the P--V diagrams. 
In Case 1, the moment 1 map exhibits a clear velocity gradient along the outflow direction but no velocity gradient perpendicular to the outflow direction (Fig.~\ref{case} (a)).
The P--V diagrams perpendicular to the outflow direction with the SMA compact and subcompact configurations (Fig.~\ref{case} (b)) and with the subcompact configuration only (Fig.~\ref{case} (c)) do not show any velocity gradient as expected. 
In Case 2 there is a clear velocity gradient perpendicular to the outflow direction but no velocity gradient along the outflow direction (Fig.~\ref{case} (d)). 
There are clear peak offsets between the blueshifted and redshifted emission in the P--V diagrams perpendicular to the outflow direction (Fig.~\ref{case} (e) \& (f)). 
In Case 3, the moment 1 map shows that the direction of the velocity gradient is from northeast to southwest (Fig.~\ref{case} (g)), 
and hence there are velocity gradients both along and perpendicular to the outflow direction (Fig.~\ref{case} (h) \& (i)).

Figure \ref{simvr} (a) and (b) show the measured rotational velocities as a function of radius derived from the simulated P--V diagrams in Case 2 and 3, respectively. 
In Case 2 (pure rotation case), the measured power-law index of the rotational profile is consistent with the input value within the 1$\sigma$ error, 
and the measured rotational velocity at a radius of 100 AU is consistent with the input value within 10\%. 
We found that over the parameter ranges we surveyed, in the pure rotation cases the measured power-law indices are consistent with the input values within 2$\sigma$ uncertainty, 
and that the rotational velocities at a radius of 100 AU within 30\%. 
On the other hand, Figure 14 (b) shows that in Case 3 the measured power-law index of the rotational profile ($p = -0.76 \pm 0.1$) is larger than the input value ($p = -1$). 
We tested several cases of the rotating and infalling envelopes and found that in these cases our method could produce artificially shallower rotational power-law indices than the genuine rotational power-law indices. 
This is because the synthetic vectors of the rotating and infalling motions can contribute to the line-of-sight velocities along the envelope major axes, 
and the P--V diagrams along the major axes do not reflect only the rotational motions but also the infalling motions. 
The magnitude of the distortion of the rotational power-law indices depends on the flatnesses of the envelopes, inclinations, ratios between the infalling and rotating motions, and the observational angular resolutions. 
For the parameters we have studied, cases of higher resolutions, flatter envelopes, higher ratios of the rotational to infalling motions, higher inclination angles (closer to the edge-on case), provide better results.

\begin{figure}
\epsscale{0.75}
\plotone{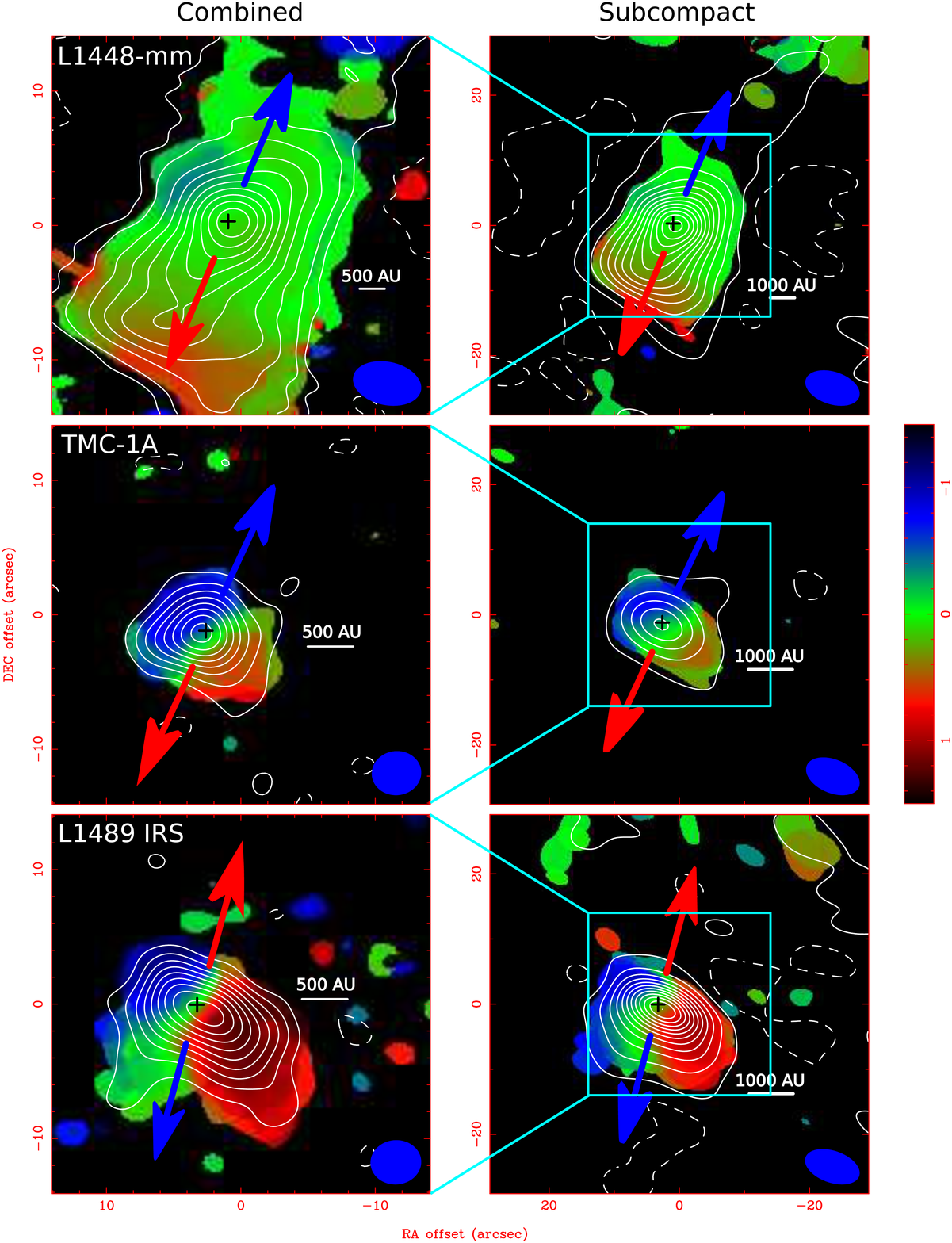} \\
Fig.~\ref{moment}.--- Continued.
\end{figure}

\begin{figure}
\epsscale{0.97}
\plotone{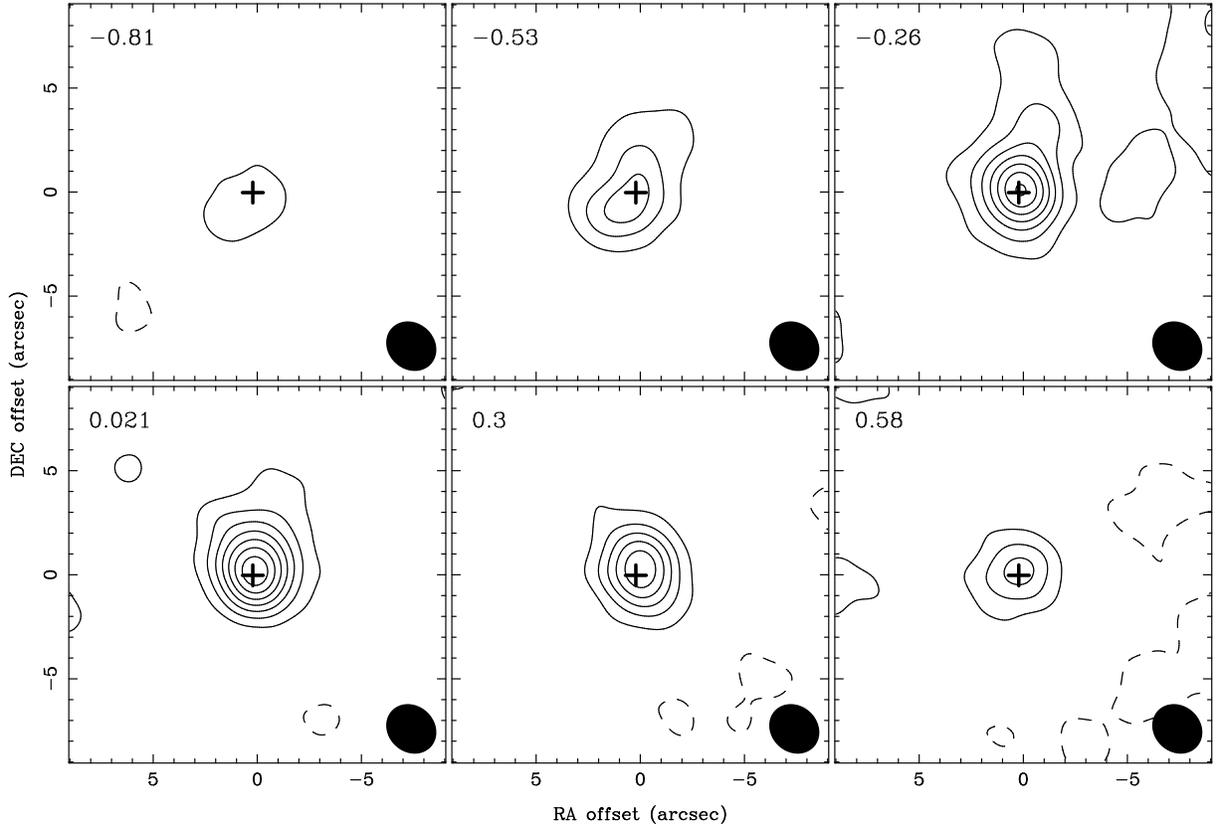}
\caption{Velocity channel maps of the C$^{18}$O (2--1) emission in IRAS 4B made with the combined data. Crosses and filled ellipses show the protostellar position and  the beam size, respectively. The velocity at each channel is shown at the upper-left corner in each panel. Contour levels start from 3$\sigma$ in steps of 4$\sigma$. The noise level and the beam size are summarized in Table \ref{resolution}.}\label{iras4B}
\end{figure}

\begin{figure}
\epsscale{0.97}
\plotone{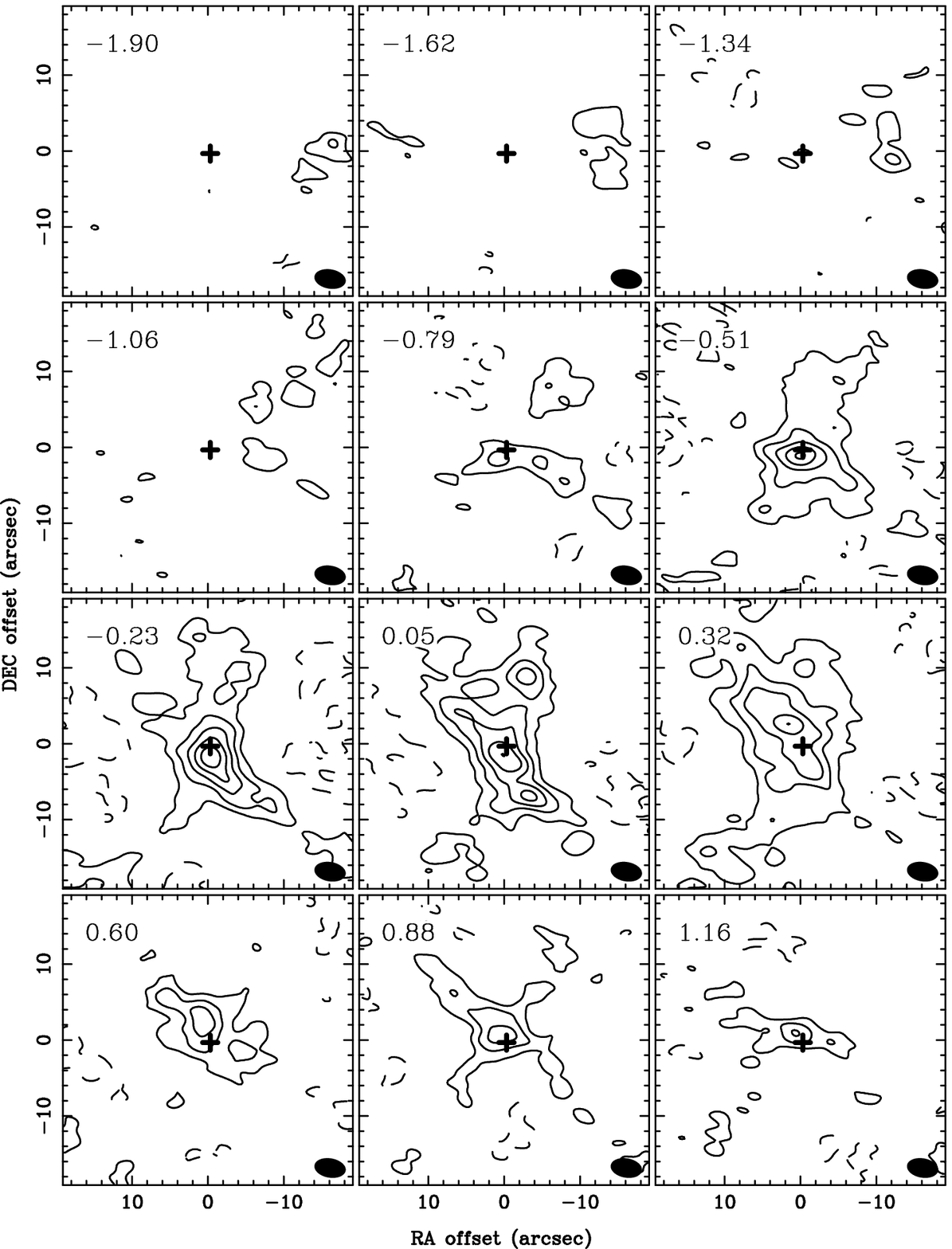}
\caption{Same as Fig.~\ref{iras4B} but for L1527 IRS. Contour levels start from 3$\sigma$ in steps of 3$\sigma$.}\label{L1527}
\end{figure}

\begin{figure}
\epsscale{0.97}
\plotone{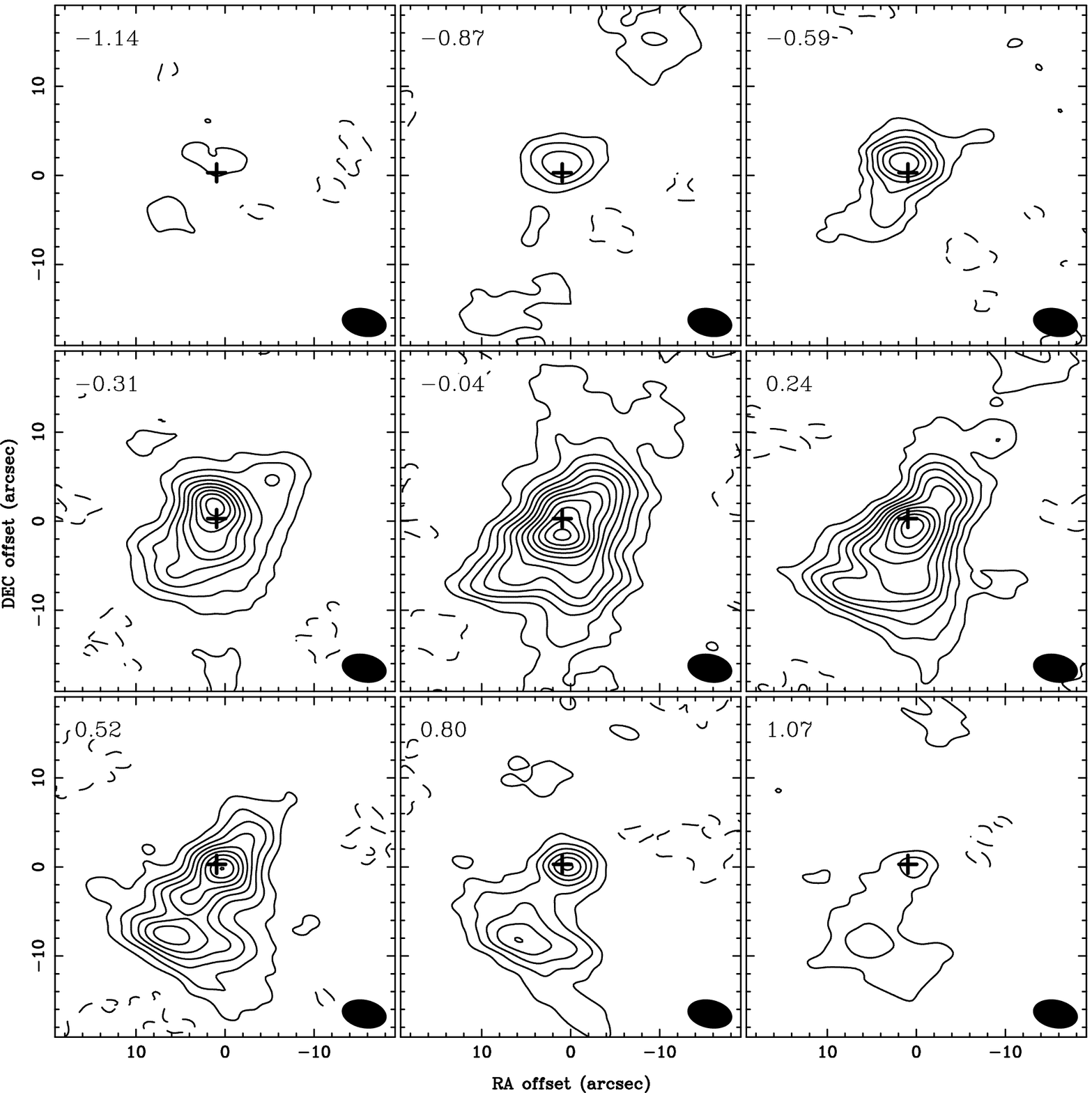}
\caption{Same as Fig.~\ref{iras4B} but for L1448-mm. Contour levels start from 3$\sigma$ in steps of 3$\sigma$.}\label{L1448}
\end{figure}

\begin{figure}
\epsscale{0.97}
\plotone{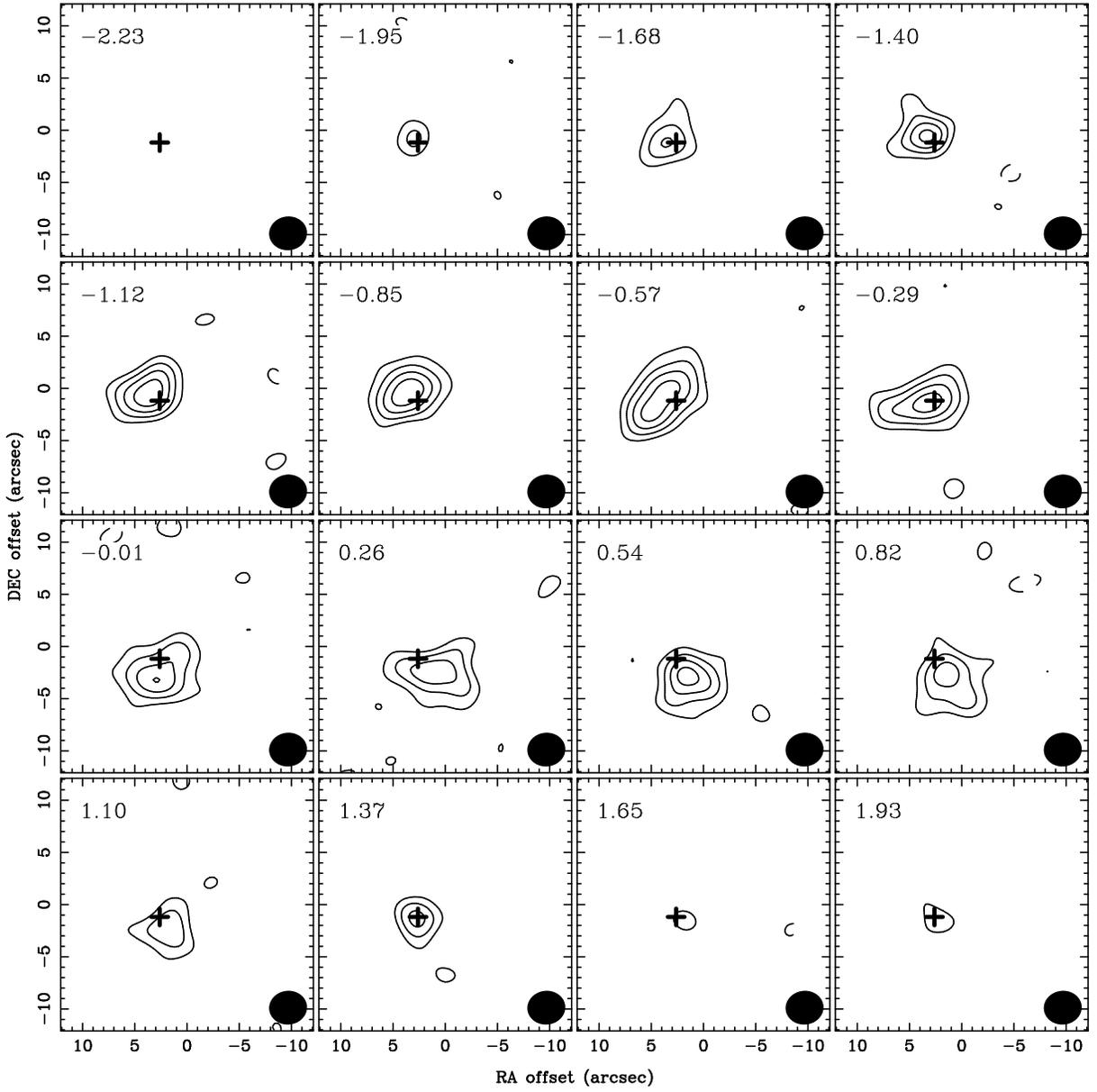}
\caption{Same as Fig.~\ref{iras4B} but for TMC-1A. Contour levels start from 3$\sigma$ in steps of 2$\sigma$.}\label{TMC1A}
\end{figure}

\begin{figure}
\epsscale{0.97}
\plotone{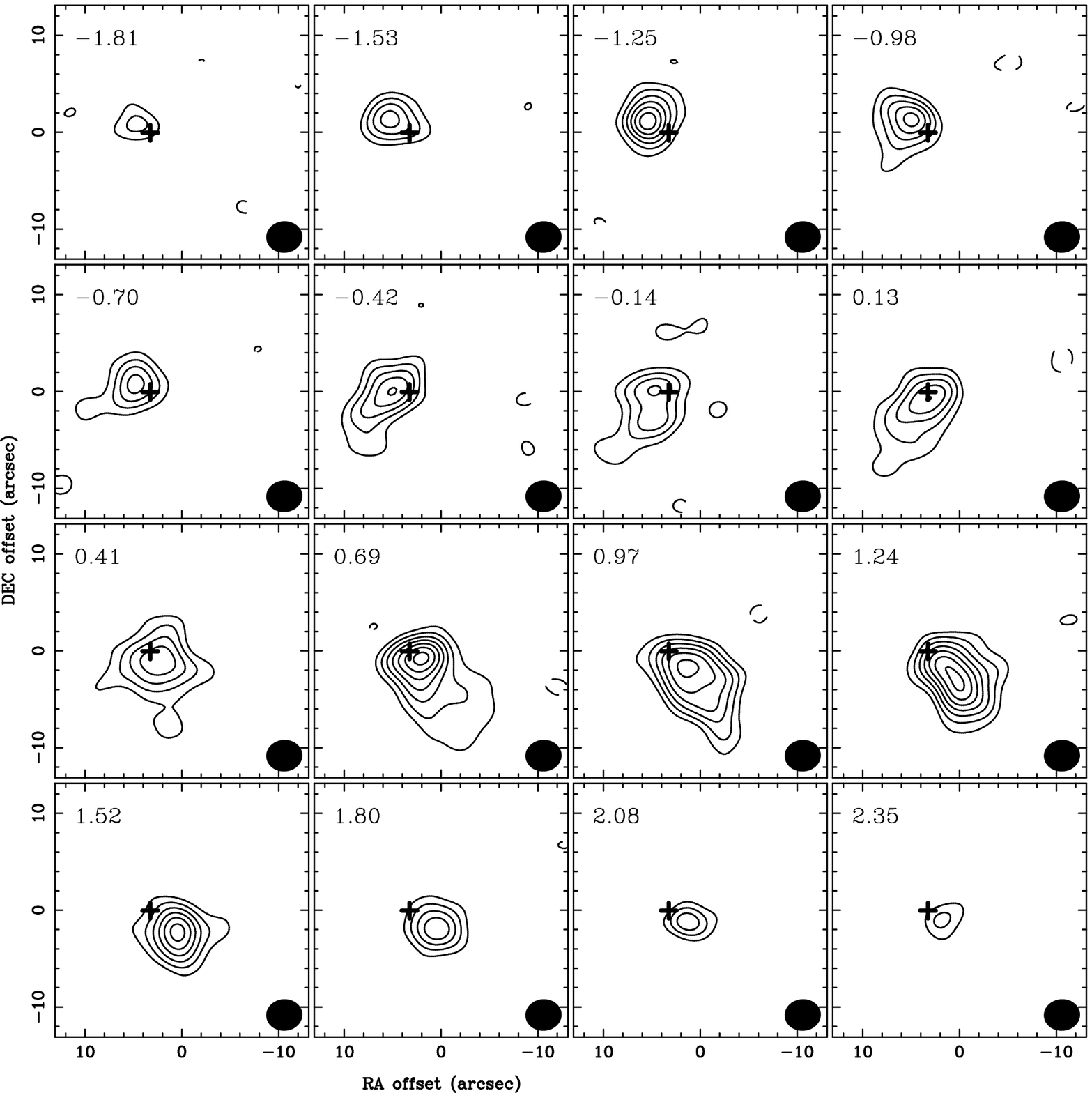}
\caption{Same as Fig.~\ref{iras4B} but for L1489 IRS. Contour levels start from 3$\sigma$ in steps of 2$\sigma$.}\label{L1489}
\end{figure}

\begin{figure}
\plotone{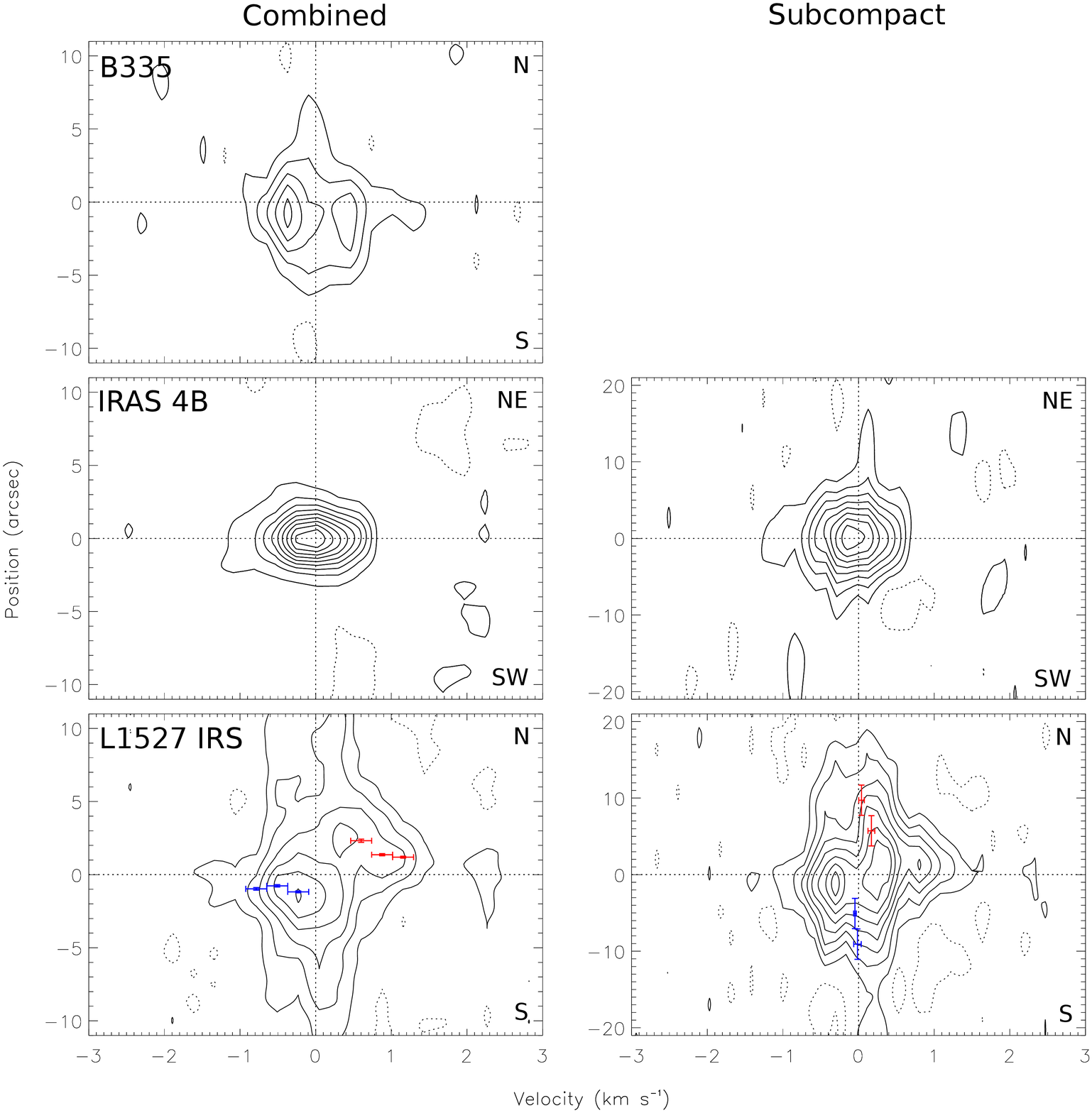}
\caption{P--V diagrams perpendicular to the outflow directions in the C$^{18}$O (2--1) emission of B335, IRAS 4B, L1527 IRS, L1448-mm, TMC-1A, and L1489 IRS. The diagrams in the left and right columns are produced with the combined and subcompact data, respectively. Note that the spatial scales of the combined and subcompact P--V diagrams are different. Horizontal and vertical lines in each panel denote the protostellar position and the systemic velocity of each source, respectively. Blue and red data points with error bars show the measured rotational velocities as a function of radius at the blueshifted and redshifted velocities, respectively. Contour levels start from 3$\sigma$ in steps of 3$\sigma$, where the noise levels are summarized in Table \ref{resolution}.}\label{pv}
\end{figure}

\begin{figure}
\plotone{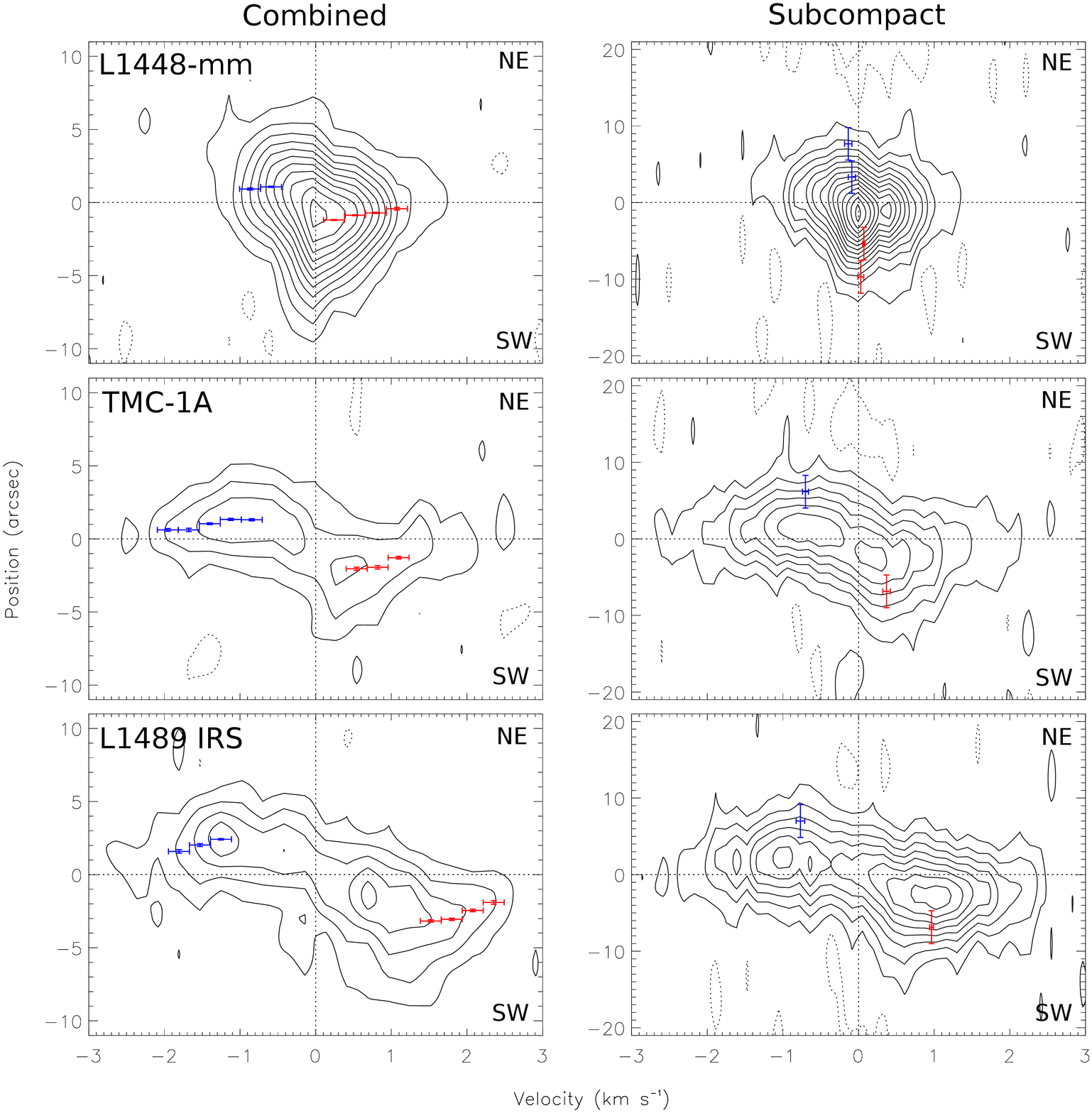}
Fig.~\ref{pv}.--- Continued.
\end{figure}

\begin{figure}
\plotone{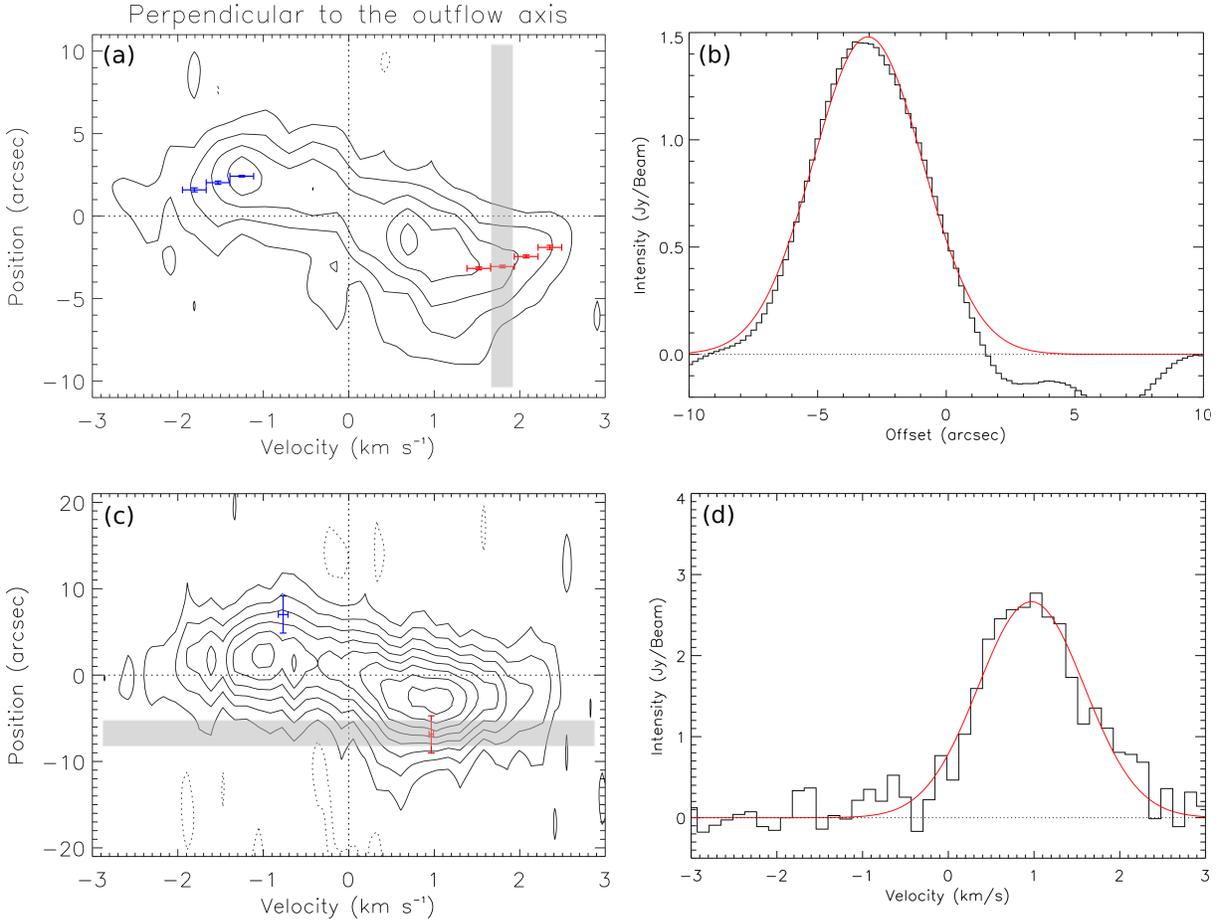}
\caption{Schematic representation of our method to derive rotational velocities as a function of radius. (a) P--V diagram of L1489 IRS in the C$^{18}$O (2--1) emission perpendicular to the outflow direction produced with the combined data (same as Figure \ref{pv}). A hatched line represents the region where a Gaussian fitting to the intensity distribution at a given velocity channel is performed. (b) Intensity distribution (black histogram) along the hatched line and the result of the Gaussian fitting (red curve). Here the centroid position of the Gaussian function is regarded as the rotational radius at that rotational velocity. (c) P--V diagram of L1489 IRS perpendicular to the outflow direction produced with the subcompact data (same as Figure \ref{pv}). A hatched line denotes the region where a Gaussian fitting to the spectrum at a given position is made. (d) Spectrum (black histogram) in the hatched region and the result of the Gaussian fitting (red curve). The centroid velocity of the Gaussian function is regarded as the rotational velocity at that rotational radius.}\label{demo}
\end{figure}

\begin{figure}
\plotone{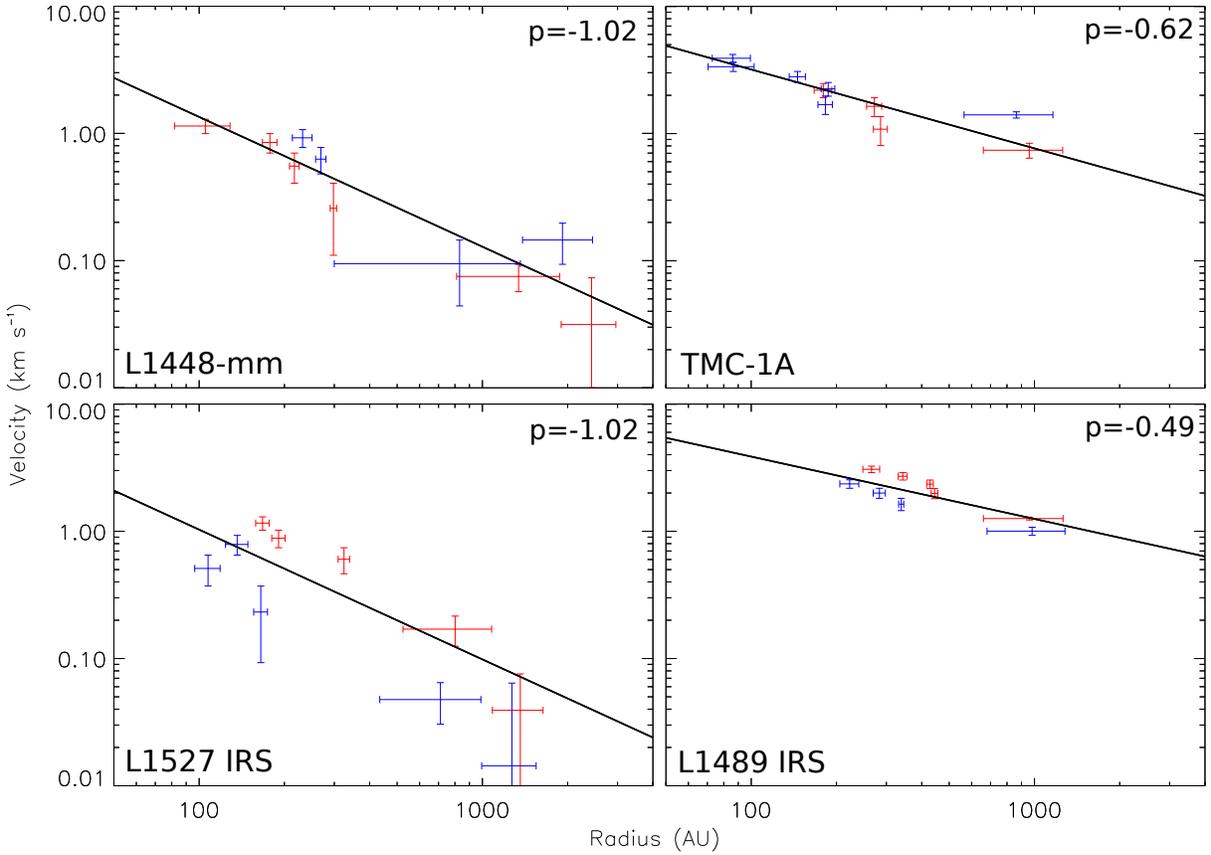}
\caption{Plots of the derived rotational velocities as a function of radius in L1527 IRS, L1448-mm, TMC-1A, and L1489 IRS. Blue and red data points are taken from Figure \ref{pv} after the correction of the inclination angles. Solid lines show the best-fit power-law rotational profiles ($v \propto r^{p}$). The best-fit power-law index is shown in the upper-right corner of each panel.}\label{vr}
\end{figure}

\begin{figure}
\plotone{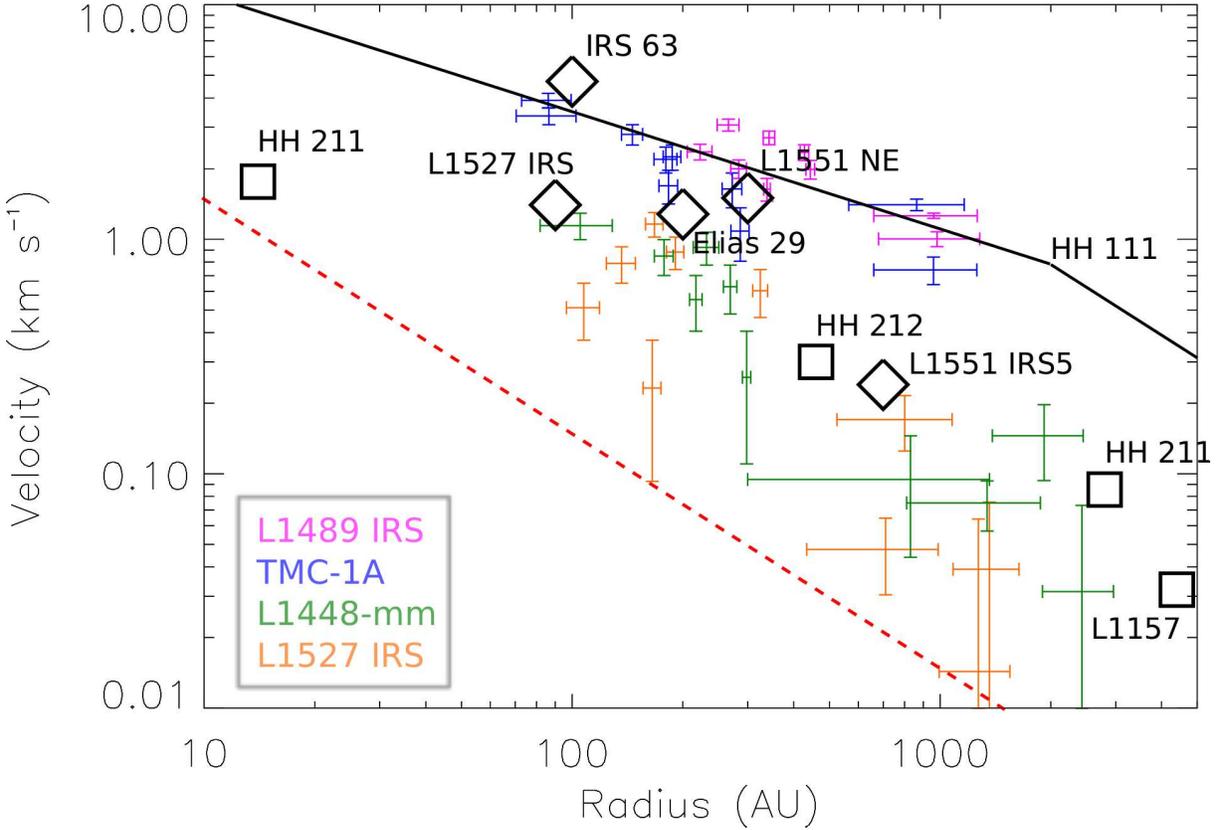}
\caption{Plots of the rotational velocities as a function of radius in L1527 IRS, L1448-mm, TMC-1A, and L1489 IRS, and comparison with those of other Class 0 and I protostars taken from the literatures. The data points of L1527 IRS, L1448-mm, TMC-1A, and L1489 IRS are shown in orange, green, blue, and magenta colors, respectively. A red dotted line shows the detection limit of our SMA observations (Yen et al.~2010), and rotational velocities around B335 and IRAS 4B are likely below this line. A black solid line presents the rotational profile of HH 111, a Class I protostellar source (Lee 2010). Open squares and diamonds show rotational velocities at the representative radii in Class 0 and I protostellar sources, respectively (Momose et al.~1998; Belloche et al.~2002; Lee et al.~2006, 2009; Lommen et al.~2008; Chiang et al.~2010; Tanner \& Arce 2011; Takakuwa et al.~2012; Tobin et al.~2012a).}\label{vrdata}
\end{figure}

\begin{figure}
\plotone{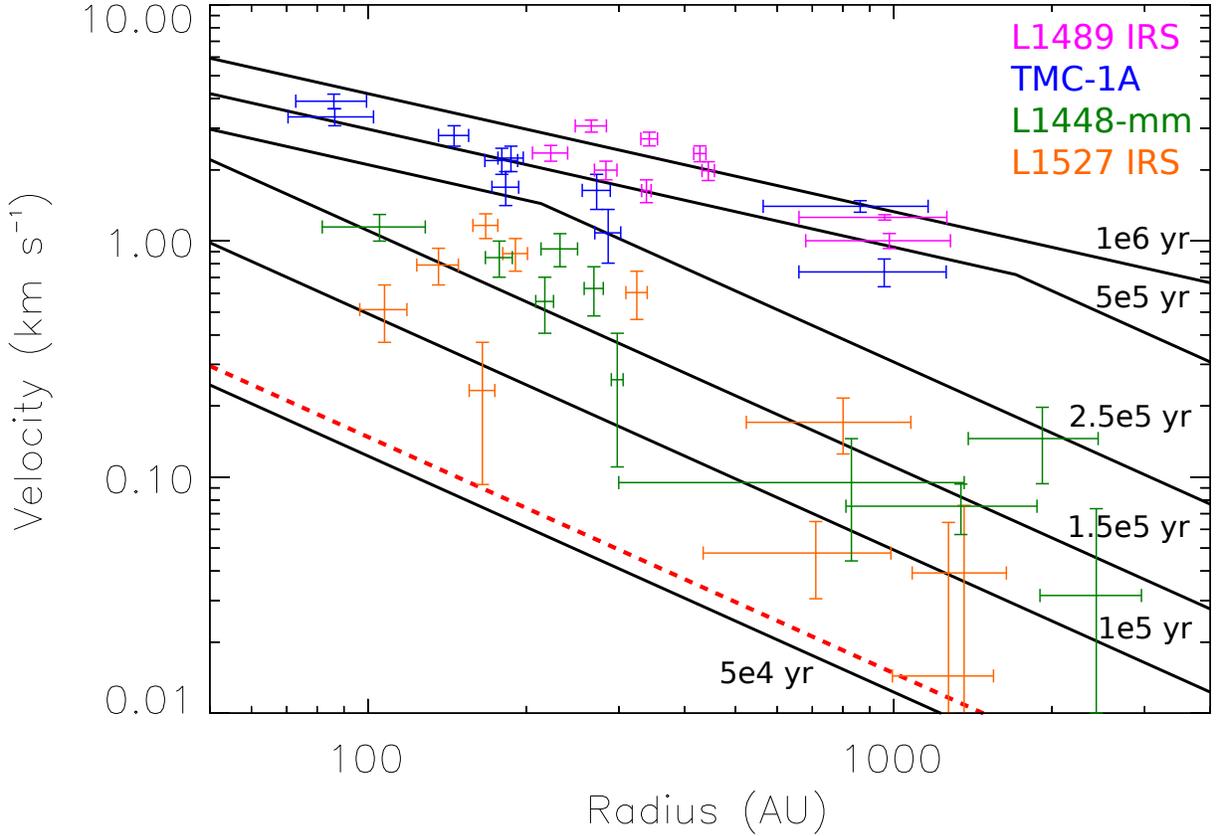}
\caption{Same as Figure \ref{vrdata} but for comparison with the expected rotational profiles calculated from our analytical model (see texts for details). Solid lines show the expected rotational profiles at different evolutionary stages of the collapse of a dense core. The breaking points in the expected rotational profiles at $t = 2.5 \times 10^{5}$ and $5 \times 10^{5}$ yr, where the power-law indices change from $-0.5$ to $-1$, present the radii of the Keplerian disks at those evolutionary stages. At $t < 1.5 \times 10^{5}$ yr, the radii of the Keplerian disks are smaller than 45 AU, and the breaking points are not seen in the plots.}\label{rdisk}
\end{figure}

\begin{figure}
\plotone{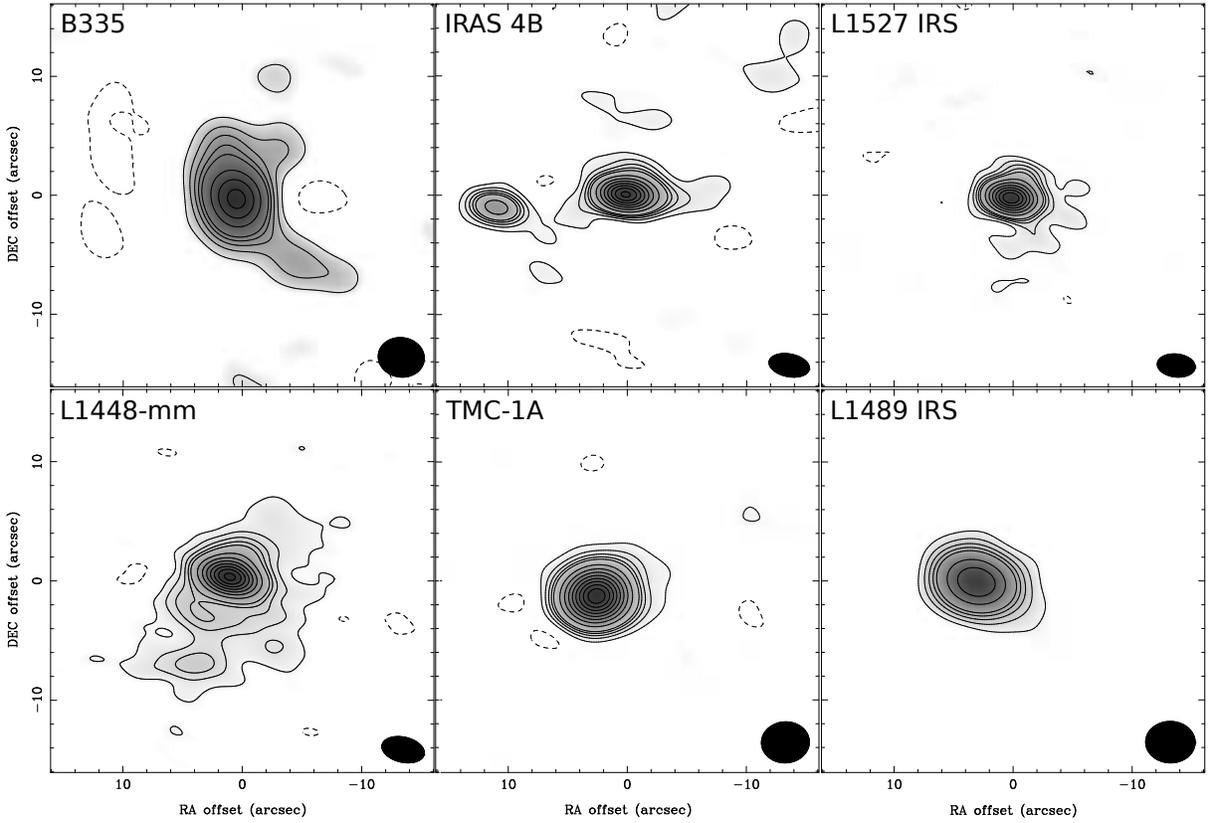}
\caption{1.3 mm continuum images of B335, IRAS 4B, L1527 IRS, L1448-mm, TMC-1A, and L1489 IRS, made with the combined SMA data. A filled ellipse at the bottom-right corner in each panel presents the beam size. Contour levels are 3$\sigma$, 6$\sigma$, 9$\sigma$, 12$\sigma$, 15$\sigma$, and then in steps of 10$\sigma$. Table \ref{contable} summarizes the 1$\sigma$ noise values.}\label{continuum}
\end{figure}

\begin{figure}
\plotone{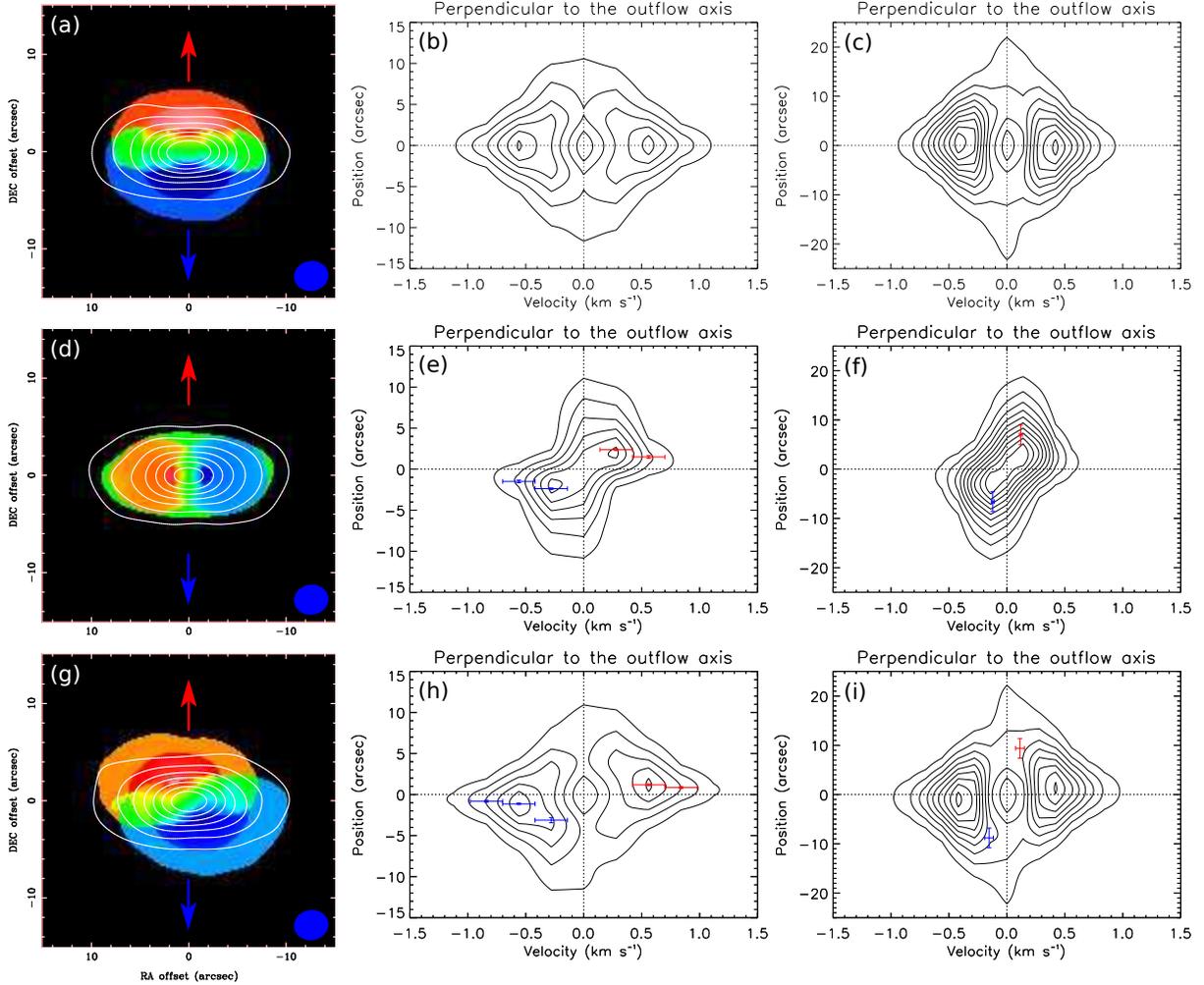}
\caption{Results of our simulations to assess the feasibility and the limitation of our method; (a) -- (c) an infalling envelope without any rotational motion; (d) -- (f) a rotating envelope without any infalling motion; and (g) -- (i) an infalling and rotating envelope. The left column shows the moment 0 maps (contours) overlaid on the moment 1 maps (color scale), produced by combining the simulated compact and subcompact data. Red and blue arrows present the outflow direction, and filled ellipses show the beam size. The middle column shows the P--V diagrams perpendicular to the outflow direction, produced with the combined data. The right column shows P--V diagrams perpendicular to the outflow direction, produced with the subcompact data. Note that the spatial scales of the combined and subcompact P--V diagrams are different. The zero position in each panel corresponds to the protostellar position. Blue and red data points with error bars show the measured rotational velocities as a function of radius.}\label{case}
\end{figure}

\begin{figure}
\plotone{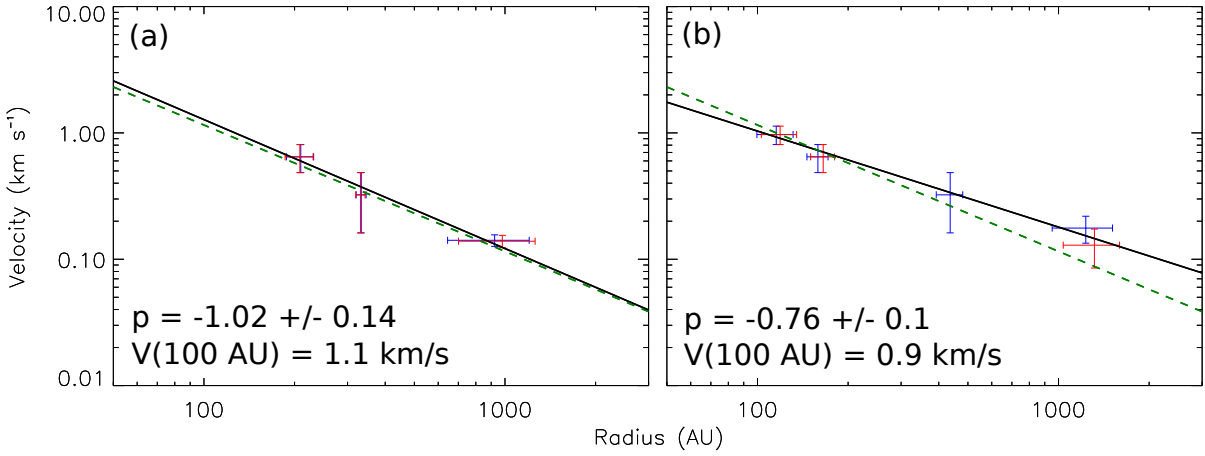}
\caption{Rotational velocities as a function of radius measured from the simulated P--V diagrams of (a) an rotating envelope without infalling motion and (b) an infalling and rotating envelope. Solid lines show the best-fit power-law rotational profiles ($v \propto r^{p}$). The best-fit power-law index and rotational velocity at a radius of 100 AU are shown at the bottom-left corner of each panel. Green dotted lines present the input rotational profile having $p = -1$ and a rotational velocity of 1 km s$^{-1}$ at a radius of 100 AU.}\label{simvr}
\end{figure}

\clearpage

\begin{deluxetable}{lcccccccccccc}
\rotate
\tablewidth{0pt}
\tablecaption{Sample of Sources}
\tablehead{\colhead{Source} & \multicolumn{2}{c}{Protostellar Position} & \colhead{Class} & \colhead{Distance} & \colhead{$L_{\rm bol}$} & \colhead{$T_{\rm bol}$} & $V_{\rm sys}$ & $i$\tablenotemark{a} & Outflow & Reference \\
\colhead{} & \multicolumn{2}{c}{(J2000)} & \colhead{} & \colhead{(pc)} & \colhead{($L_{\sun}$)} & \colhead{(K)} & \colhead{(km s$^{-1}$)} & \colhead{} & \colhead{(P.A.)} & \colhead{}
}
\startdata
B335 & 19$^{h}$37$^{m}$00$\fs$93 & 07\arcdeg34\arcmin09$\farcs$8 & 0 & 150 & 1.5 & 31 & 8.3 & 80\degr & 90\degr & 1,2,3,4,5\\
NGC 1333 IRAS 4B & 03$^{h}$29$^{m}$12$\fs$01 & 31\arcdeg13\arcmin08$\farcs$1 & 0 & 250 & 1.6 & 54 & 6.7 & 77\degr & 151\degr & 1,6,7,8,9\\
L1527 IRS & 04$^{h}$39$^{m}$53$\fs$91 & 26\arcdeg03\arcmin09$\farcs$8 & 0/I & 140 & 2.8 & 59 & 5.7 & 85\degr & 90\degr & 1,10,11\\
L1448-mm & 03$^{h}$25$^{m}$38$\fs$87 & 30\arcdeg44\arcmin05$\farcs$4 & 0 & 250 & 7.5 & 69 & 5.0 & 70\degr & 157\degr & 1,6,12,13,14\\
TMC-1A & 04$^{h}$39$^{m}$35$\fs$20 & 25\arcdeg41\arcmin44$\farcs$4 & I & 140 & 2.4 & 172 & 6.6 & 20$\degr$--50$\degr$\tablenotemark{b} & 155\degr & 1,15,16,17\\
L1489 IRS & 04$^{h}$04$^{m}$42$\fs$85 & 26\arcdeg18\arcmin56$\farcs$3 & I & 140 & 3.7 & 238 & 7.2 & 36$\degr$--90$\degr$\tablenotemark{c} & 165\degr & 1,15,16,18 \\
\enddata
\tablenotetext{a}{The inclination angle ($i$) is defined as the angle between the disk plane and the plane of the sky, i.e., an inclination angle of 90$\degr$ corresponds to the edge-on geometry.}
\tablenotetext{b}{A medium value of 30$\degr$ is adopted in this paper.}
\tablenotetext{c}{A medium value of 50$\degr$ is adopted in this paper.}
\tablerefs{(1) Motte \& Andr{\'e} 2001; (2) Stutz et al.~2008; (3) Chandler \& Sargent 1993; (4) Yen et al.~2011; (5) Hirano et al.~1988; (6) Enoch et al.~2006; (7) Enoch et al.~2009b; (8) Volgenau et al.~2006; (9) Marvel et al.~2008; (10) Tobin et al.~2008; (11) Ohashi et al.~1997a; (12) Tobin et al.~2007; (13) Curiel et al.~1999; (14) Girart \& Acord 2001; (15) Furlan et al.~2008; (16) Hogerheijde et al.~1998; (17) Chandler et al.~1996; (18) Brinch et al.~2007a, b.}
\label{sample}
\end{deluxetable}

\begin{deluxetable}{lccccc}
\tablewidth{0pt}
\tablecaption{Summary of Observations}
\tablehead{ & \multicolumn{5}{c}{Subcompact Configuration}}
\startdata
 Source & IRAS 4B & L1527 IRS & L1448-mm & TMC-1A & L1489 IRS \\ 
 \hline 
 Observing Date & 2012 Jan  7 & 2012 Jan  8 & 2012 Jan  7 & 2012 Jan  8 & 2012 Jan  7\\
 Pointing Center (J2000) & 03$^{h}$29$^{m}$12$\fs$00 & 04$^{h}$39$^{m}$53$\fs$90 & 03$^{h}$25$^{m}$38$\fs$80 & 04$^{h}$39$^{m}$35$\fs$01 & 04$^{h}$04$^{m}$42$\fs$95 \\
 & 31\arcdeg13\arcmin08$\farcs$0 & 26\arcdeg03\arcmin10$\farcs$0 & 30\arcdeg44\arcmin05$\farcs$0 &  25\arcdeg41\arcmin45$\farcs$5 & 26\arcdeg18\arcmin56$\farcs$3 \\ 
 225 GHz Opacity & \multicolumn{5}{c}{0.03 -- 0.12} \\
 System Temperature & \multicolumn{5}{c}{70 -- 130 K} \\
 Bandpass Calibrator & \multicolumn{5}{c}{3c84 and 3c279} \\
 Flux Calibrator & \multicolumn{5}{c}{Uranus} \\
 Gain Calibrator & \multicolumn{5}{c}{3c111 (2.1 Jy) and 3c84 (8.8 Jy)}\\ 
 $uv$ Coverage & \multicolumn{5}{c}{3 -- 34 $k\lambda$} \\
 \hline \hline \\
 & \multicolumn{2}{c}{Compact Configuration} & & \multicolumn{2}{c}{Extended Configuration}\\
 \hline
 Source & TMC-1A & L1489 IRS & & \multicolumn{2}{c}{IRAS 4B}\\
 \hline
 Observing Date & \multicolumn{2}{c}{2012 Oct 28} & & \multicolumn{2}{c}{2006 Jan 17} \\
 Pointing Center (J2000) &  04$^{h}$39$^{m}$35$\fs$01 & 04$^{h}$04$^{m}$42$\fs$95 & & \multicolumn{2}{c}{03$^{h}$29$^{m}$12$\fs$00} \\
 & 25\arcdeg41\arcmin45$\farcs$5 & 26\arcdeg18\arcmin56$\farcs$3 & & \multicolumn{2}{c}{31\arcdeg13\arcmin08$\farcs$0} \\ 
 225 GHz Opacity & \multicolumn{2}{c}{0.1 -- 0.2} & & \multicolumn{2}{c}{$\sim$ 0.07} \\
 System Temperature & \multicolumn{2}{c}{100 -- 300 K} & & \multicolumn{2}{c}{80 -- 250 K} \\ 
 Bandpass Calibrator & \multicolumn{2}{c}{3c279 and 3c454.3} & & \multicolumn{2}{c}{3c273} \\
 Flux Calibrator & \multicolumn{2}{c}{Uranus} & & \multicolumn{2}{c}{Callisto} \\
 Gain Calibrator & \multicolumn{2}{c}{3c111 (1.9 Jy)} & & \multicolumn{2}{c}{3c84 (2.8 Jy) and 3c111 (3.3 Jy)} \\ 
 $uv$ Coverage & \multicolumn{2}{c}{8 -- 57 $k\lambda$} & & \multicolumn{2}{c}{17 -- 134 $k\lambda$} 
\enddata
\label{obsummary}
\end{deluxetable}

\begin{deluxetable}{lclcclc}
\rotate
\tablewidth{0pt}
\tablecaption{Resolutions and Noise Levels of the C$^{18}$O (2--1) Images}
\tablehead{ 
\colhead{} & \colhead{} & \multicolumn{2}{c}{Subcompact Image} & & \multicolumn{2}{c}{Combined Image} \\
\cline{3-4} \cline{6-7} \\
\colhead{Source} & \colhead{$u$--$v$ coverage ($k\lambda$)} & \colhead{Beam (P.A.)} & \colhead{RMS (Jy Beam$^{-1}$)} & & \colhead{Beam (P.A.)} & \colhead{RMS (Jy Beam$^{-1}$)} } 
\startdata
B335 & 6 -- 54 & \nodata & \nodata & & 3\farcs7 $\times$ 3\farcs2 (87\degr) & 0.20  \\
IRAS 4B & 3 -- 134 & 8\farcs5 $\times$ 4\farcs8 (73\degr) & 0.19 & & 2\farcs6 $\times$ 2\farcs2 (47\degr) & 0.09 \\
L1527 IRS & 4 -- 102 & 8\farcs5 $\times$ 5\farcs1 (68\degr) & 0.17 & & 4\farcs2 $\times$ 2\farcs5 (80\degr) & 0.10 \\
L1448-mm & 3 -- 102 & 8\farcs5 $\times$ 4\farcs9 (72\degr) & 0.18 & & 5\farcs1 $\times$ 3\farcs2 (78\degr) & 0.11\\
TMC-1A & 4 -- 57 & 8\farcs6 $\times$ 5\farcs1 (67\degr) & 0.17 & & 3\farcs6 $\times$ 3\farcs2 (-84\degr) & 0.16 \\
L1489 IRS & 3 -- 57 & 8\farcs5 $\times$ 4\farcs9 (72\degr) & 0.18 & & 3\farcs7 $\times$ 3\farcs3 (78\degr) & 0.16 \\
\hline \\
\multicolumn{2}{c}{Velocity Resolution} & \multicolumn{2}{c}{0.14 km s$^{-1}$} &  & \multicolumn{2}{c}{0.28 km s$^{-1}$}
\enddata
\label{resolution}
\end{deluxetable}

\begin{deluxetable}{lccc}
\tablewidth{0pt}
\tablecaption{Power-law Indices of the Rotational Profiles and the Estimated Protostellar Mass}
\tablehead{ \colhead{Source} & \colhead{$p$} & \colhead{Protostellar Mass ($M_{\sun}$)}}
\startdata
B335 & \nodata & \nodata \\
IRAS 4B & \nodata & \nodata \\
L1527 IRS & $-1.02\pm0.17$ & \nodata \\
L1448-mm & $-1.02\pm0.14$ & \nodata \\
TMC-1A & $-0.62\pm0.07$ & $1.1\pm0.1$\tablenotemark{a} \\
L1489 IRS & $-0.49\pm0.05$ & $1.8\pm0.2$\tablenotemark{a} \\
\enddata
\tablenotetext{a}{The estimated protostellar mass is sensitive to the inclination angle by $1/\sin^2(i)$.}
\label{power}
\end{deluxetable}

\begin{deluxetable}{llccl}
\rotate
\tablewidth{0pt}
\tablecaption{Summary of the 1.3 mm Continuum Observations}
\tablehead{ 
\colhead{Source} & \colhead{Beam (P.A.)} & \colhead{RMS (mJy Beam$^{-1}$)} & \colhead{Total Flux (Jy)} & \colhead{Deconvolved Size (P.A.)}} 
\startdata
B335 &  3\farcs9 $\times$ 3\farcs3 (82\degr) & 2 & 0.18 & 4\farcs9 $\times$ 2\farcs3 (13\degr) \\
IRAS 4B &  3\farcs4 $\times$ 1\farcs9 (79\degr) & 9 & 1.11 & 2\farcs3 $\times$ 1\farcs2 (-77\degr) \\
L1527 IRS & 3\farcs2 $\times$ 1\farcs9 (84\degr) & 2 & 0.21 & 1\farcs4 $\times$ 1\farcs2 (-8\degr) \\
L1448-mm & 3\farcs6 $\times$ 2\farcs1 (77\degr) & 2 & 0.26 & 2\farcs1 $\times$ 2\farcs0 (-79\degr) \\
TMC-1A &  4\farcs0 $\times$ 3\farcs5 (-87\degr) & 2 & 0.23 & 1\farcs6 $\times$ 1\farcs2 (-71\degr) \\
L1489 IRS & 4\farcs2 $\times$ 3\farcs5 (89\degr) & 1 & 0.07 & 3\farcs7 $\times$ 2\farcs0 (67\degr) \\
\enddata
\label{contable}
\end{deluxetable}

\end{document}